\documentclass[12pt,a4paper]{article}

\usepackage[utf8]{inputenc}
\usepackage[T1]{fontenc}
\usepackage{lmodern}
\usepackage{amsmath,amssymb,amsthm}
\usepackage{geometry}
\usepackage{graphicx}
\usepackage{booktabs}
\usepackage{tabularx}
\usepackage{array}
\usepackage{multirow}
\usepackage{float}
\usepackage{caption}
\usepackage{subcaption}
\usepackage{enumitem}
\usepackage{xcolor}
\usepackage{lineno}
\usepackage[numbers,sort&compress]{natbib}
\usepackage{xr-hyper}
\usepackage[hidelinks]{hyperref}
\usepackage{tikz}
\usetikzlibrary{shapes.geometric,arrows.meta,positioning}

\graphicspath{{../figures/}}

\newcommand{\R}{\mathbb{R}}
\newcommand{\E}{\mathbb{E}}

\newcommand{\F}{\mathcal{F}}
\newcommand{\norm}[1]{\lVert#1\rVert}
\newcommand{\given}{\,|\,}
\newcommand{\LRM}{\mathcal{L}_{\mathrm{RM}}}
\newcommand{\Ltask}{\mathcal{L}_{\mathrm{task}}}
\newcommand{\Ltot}{\mathcal{L}_{\mathrm{total}}}
\newcommand{\RMRNN}{\textsc{RMRNN}}

\graphicspath{{figures/}{./}}
\begin{document}

\title{\textbf{Coupling Precipitation Forecasting and Early Warning
with Reverse-Martingale Recurrent Neural Networks}}

\author{Foo Hui-Mean$^{\,1}$  ~~   Yuan-chin Ivan Chang$^{\,1,*}$ \\[4pt]
\small $^{1}$Institute of Statistical Science, Academia Sinica, Taipei, Taiwan \\
\small $^{*}$Corresponding author: \texttt{ycchang@stat.sinica.edu.tw}}

\date{\today}
\maketitle

\begin{abstract}
Precipitation forecasts are judged by accuracy, but the decisions they support
--- when to restrict water, when to warn of drought --- turn on noticing when a
local regime is becoming abnormal, which forecast scores alone do not reveal. We
ask whether one recurrent model can do both with little or no loss in forecast
skill. We add a backward-coherence (reverse-martingale) penalty that keeps the
network's hidden state smooth when read backward in time; the size of the
resulting reconstruction defect becomes an online warning signal, monitored by a
sequential change-point detector. The design is deliberately conservative. On
real daily station data from four contrasting climates --- monsoonal Taiwan,
semi-arid Texas, temperate Germany, and Mediterranean Anatolia (Turkey) --- the model matches a standard
network's forecast skill everywhere, and makes the hidden state markedly steadier
in every region. The novelty is the added information: on these real droughts the
signal can alarm well ahead of the operational SPI-3 index, giving lead that
neither the forecast nor the index provides.
This benefit is not uniform across the four regions --- large in one, partial in
two others, and near-absent in the fourth. We offer the hydroclimatic character of drought
onset, whether it precedes or merely coincides with the rainfall deficit, as a
plausible explanation to be tested in future work, supported by a controlled
synthetic study with known onset times. The contribution is thus a new and
conservative way to read precipitation records: no loss in forecast skill, a
steadier model, and an early-warning signal beyond the standard index.
\end{abstract}


\section*{Highlights}
\begin{itemize}\setlength{\itemsep}{1pt}
  \item A backward-coherence penalty turns a precipitation RNN into a warning sensor
  \item Forecast skill is preserved across four contrasting real climate regions
  \item The hidden-state defect can alarm drought onset ahead of the SPI-3 index
  \item Lead is region-dependent: largest where onset precedes the rainfall deficit
  \item A synthetic study with known onset times explains where the advantage holds
\end{itemize}

\section{Introduction}
\label{sec:intro}

Operational hydrometeorology increasingly needs precipitation products that serve not only as forecasts but also as warnings. Reservoir operations, irrigation, urban drainage, drought mitigation, and flash-flood response all depend on detecting damaging regimes early enough for action \citep{ravuri2021,espeholt2022}. This is especially difficult at small spatial scales, where local precipitation is shaped by terrain, land--sea contrast, convective organization, and rapidly evolving moisture transport \citep{roe2005}. As a result, a model may achieve reasonable average forecast scores while still issuing warnings that are late, noisy, or poorly calibrated.

Deep recurrent models, including LSTM \citep{hochreiter1997}, GRU \citep{cho2014}, ConvLSTM \citep{shi2015}, and PredRNN \citep{wang2017predrnn}, are well suited to extracting temporal structure from gauge, satellite, radar, and reanalysis sequences. For operational warning, however, the forecast value alone is often insufficient. Thresholding a predicted rain rate, accumulated precipitation total, or drought index treats warning as post-processing and ignores information in the evolving hidden state. This separation is especially limiting when drought, monsoon transition, typhoon rainbands, or flash-flood-producing regimes begin to emerge before local precipitation thresholds are crossed \citep{gama2014,milly2008}.

Operational drought monitoring and prediction is itself a mature field, spanning multi-source indicators, statistical and dynamical prediction, uncertainty communication, and user-oriented early warning at regional to global scales \citep{hao2017}. Within that broad agenda we pursue a narrower, complementary question: given a recurrent precipitation forecasting model, how can its evolving internal state be used to produce a stable, calibrated drought-warning signal? To that end, this paper develops a forecast-preserving warning framework for local, station-scale precipitation. The key idea is to regularize the recurrent hidden state so that ordinary hydroclimatic evolution remains coherent when read backward through a learned one-step projector. The resulting backward-coherence residual is computed online and passed to a sequential change-point detector, allowing the same hidden trajectory to produce both the precipitation forecast and a calibrated drought--flood warning signal. The aim is deliberately conservative: rather than improving ordinary forecast accuracy, the method seeks to add early-warning capability with little or no loss in forecast skill. We therefore ask whether reverse-martingale (RM) regularization can preserve GRU-level forecast performance while turning the hidden-state trajectory into a useful warning diagnostic, and in which hydroclimatic settings it provides meaningful lead time.

\textbf{Main contributions.}
This paper introduces a forecast-preserving warning workflow for local precipitation. A recurrent model produces the usual probabilistic forecast while its hidden-state trajectory also yields an online RM defect,
\(
d_t=\norm{h_{t-1}-g_\phi(h_t)},
\)
computed after the observation at time $t$ is assimilated. The standardized defect drives a one-sided CUSUM detector calibrated by ARL$_0$ on held-out climatology.
We evaluate the framework on four real daily station networks representing distinct precipitation climates: monsoonal, mountainous Taiwan; semi-arid Texas Hill Country; temperate, maritime-influenced Germany; and Mediterranean-continental Turkey (Section~\ref{sec:realeval}). Across all four regions, the reverse-martingale recurrent neural network (\RMRNN{}) preserves GRU-level forecast and heavy-rain warning skill while reducing hidden-state path instability by $43$--$55\%$ in $Q_{\mathrm{path}}$. The novelty is in what the same model \emph{additionally} provides: an early-warning signal, read directly from the hidden state, that on these real droughts can alarm well ahead of the operational SPI-3 index --- information that neither the forecast nor the standard index supplies. This added lead is not uniform across the four regions --- large in semi-arid Texas, smaller but still positive in temperate Germany, marginal in Mediterranean-continental Turkey, and near zero in monsoonal Taiwan. We do \emph{not} treat this variation as an established effect of landform: establishing that would require many more regions and is beyond the scope of this paper. We instead report it as an observation and offer the hydroclimatic character of drought onset --- whether it precedes or merely coincides with the rainfall deficit --- as a plausible explanation to be tested in future work. A controlled synthetic study with known injected onsets supports this reading: the defect leads for slow-onset drought but not for sharp heavy-rain onset, which is already visible in precipitation itself (Section~\ref{sec:risk}).

The rest of the paper is organized as follows. Section~\ref{sec:method} presents the coupled forecasting-and-warning workflow, with RM details deferred to Appendix~\ref{app:rm-details}. Section~\ref{sec:realeval} gives the real-data evaluation. Section~\ref{sec:data} presents the controlled synthetic experiments, including false-alarm calibration and the multi-variable defect. Sections~\ref{sec:discuss} and \ref{sec:concl} discuss implications, limitations, and operational extensions. Additional synthetic results and full data-acquisition and preprocessing details are provided in the online supplement.

\section{Forecasting and Warning Workflow}
\label{sec:method}

This section presents the operational version of the proposed method. The central idea is simple: the recurrent model produces a standard precipitation forecast, while its hidden-state trajectory is monitored for unusual departures from normal local evolution. The main text focuses on the forecasting-and-warning workflow; formal reverse-martingale notation, loss functions, and implementation details are collected in Appendix~\ref{app:rm-details}.

\subsection{Local inputs and forecast target}

At each issue time, the model uses recent meteorological observations from a small neighborhood around the target basin, station, or grid cell. Inputs vary by dataset and may include precipitation, temperature, humidity, vorticity, soil moisture, and low-level winds. Neighborhood radius is tuned for the Taiwan CWA network, where precipitation is strongly shaped by orographic and basin-scale effects, and fixed for the lower-density CHIRPS and GHCN-Daily applications. Full notation is provided in Appendix~\ref{app:rm-details}.

The recurrent model encodes local meteorological history into a hidden state $h_t$ and produces a probabilistic accumulated-precipitation forecast at the chosen lead time. To handle frequent zeros and a skewed positive tail, precipitation is modeled with a mixed distribution: a dry probability for zero rainfall and a log-normal distribution for positive amounts, following \citet{sloughter2007}. We use the Continuous Ranked Probability Score $\mathrm{CRPS}$ as the primary verification metric and report $\mathrm{RMSE}$ and $\mathrm{MAE}$ for comparison.

\subsection{Backward coherence as a warning diagnostic}

The reverse-martingale regularizer encourages neighboring hidden states to remain coherent during typical hydroclimatic evolution. It promotes hidden trajectories in which each subsequent state retains enough information to approximately reconstruct the previous state. Smooth local weather evolution should therefore yield small reconstruction defects, while abrupt transitions—such as drought onsets, monsoon shifts, frontal passages, or typhoon rainbands—should produce larger, measurable departures.

A key directionality applies to drought: once a sustained dry regime sets in, precipitation variability drops and atmospheric dynamics become simpler and more repetitive. The hidden state therefore evolves more predictably, and the backward projector $g_\phi$ reconstructs it more accurately — so the defect $d_t$ \emph{decreases} relative to its climatological level. The CUSUM monitors this \emph{downward} shift as a drought signal (Section~\ref{sec:srnote}). The opposite applies to the onset of heavy-rain events, which raises the defect.

This regularizer does not assume that precipitation follows a martingale, nor is it designed to minimize RMSE. Its purpose is to make the hidden state useful as a stability diagnostic while preserving forecast skill. The joint objective combines the standard precipitation loss with a backward-coherence penalty; the loss function, training schedule, and projector architecture are detailed in Appendix~\ref{app:rm-details}.

\subsection{Online residual and alarm rule}

For operational use, the warning residual must be available when an alert is issued. After assimilating the observation at valid time $t$ and updating the hidden state, we compute
\begin{equation}
d_t = \norm{h_{t-1}-g_\phi(h_t)},
\label{eq:defect}
\end{equation}
where $g_\phi$ is the learned backward projector. Because this defect uses only information available up to time $t$, we set $r_t \equiv d_t$ in the warning experiments.

\label{sec:srnote}%
We standardize the residual on pre-event climatology and monitor it with a one-sided cumulative-sum (CUSUM) sequential change-point detector \citep{moustakides1986},
\begin{equation}
S_t = \max\!\big(0,\; S_{t-1} + z_t - k\big), \qquad S_0 = 0,
\label{eq:cusum}
\end{equation}
with reference value $k=1/2$, where $z_t$ is the standardized defect. The sign of $z_t$ is chosen to match the expected direction of the regime shift: for \emph{drought}, the defect decreases under a dry regime (as explained in Section~\ref{sec:method}), so we monitor a downward shift (i.e.\ negate $z_t$ before accumulating); for \emph{heavy-rain onset}, the defect increases, so we monitor an upward shift directly. This means the CUSUM statistic $S_t$ always increases under the regime of concern, regardless of direction. The detector alarms when $S_t$ first exceeds a threshold $h$ calibrated to a target no-change average run length, $\text{ARL}_0$, by Monte Carlo simulation on held-out climatological periods rather than by asymptotic formulas alone. We use the additive CUSUM because the standardized reverse-martingale defect is strongly right-skewed and autocorrelated: it is the classical sequential detector for exactly this heavy-tailed, dependent setting, and it attains the target $\text{ARL}_0$ stably where a multiplicative recursion would be driven to premature alarms by occasional large values. All warning results in Section~\ref{sec:risk} use this detector.

\begin{figure}[ht]
\centering
\fbox{\begin{minipage}{0.92\linewidth}
\vspace{2pt}
\textbf{Algorithm 1.} \textit{Residual-driven sequential warning detector (the one-sided CUSUM of Eq.~\eqref{eq:cusum}, Section~\ref{sec:srnote}).}
\vspace{2pt}\hrule\vspace{4pt}
\textbf{Require:} Trained recurrent model and backward projector; held-out no-change climatology; target $\text{ARL}_0$; reference value $k=1/2$; shift direction for the regime ($-$ for drought, $+$ for heavy-rain onset).\\
\textbf{Ensure:} Calibrated threshold and online alarm sequence.
\begin{enumerate}[leftmargin=1.5em,itemsep=1pt,topsep=2pt]
\item Compute online hidden-state residuals $r_t=\norm{h_{t-1}-g_\phi(h_t)}$ on the held-out null climatology.
\item Estimate the null location $\hat\mu_0$ and scale $\hat\sigma_0$ (after the variance-stabilizing transform described in Appendix~\ref{app:rm-details}) and form the standardized residual $z_t=(r_t-\hat\mu_0)/\hat\sigma_0$.
\item Run the one-sided CUSUM $S_t=\max(0,\,S_{t-1}+s\,z_t-k)$ where $s$ is the regime shift direction, and select the threshold $h$ via Monte Carlo sampling on bootstrapped null segments such that the mean time to a false alarm matches the target $\text{ARL}_0$.
\item During online operation, update the hidden state $h_t$ upon receiving each new observation, compute $r_t$ and $z_t$, update $S_t$, and trigger an alarm at the first instance where $S_t \ge h$.
\end{enumerate}
\vspace{2pt}
\end{minipage}}
\end{figure}


\section{Real-Data Evaluation}
\label{sec:realeval}

We begin with real observations across multiple landforms, since the paper’s operational claims concern real precipitation. The four networks are chosen to be \emph{deliberately contrasting along two axes at once}: four distinct landform-and-weather regimes --- monsoonal and steeply orographic, semi-arid subtropical, temperate maritime, and Mediterranean-continental --- sampled at four widely separated locations on the globe (East Asia, North America, Central Europe, and the eastern Mediterranean). Spanning both climate type and geographic setting is deliberate: a property that holds across all four is unlikely to be an artefact of any single region's climate, terrain, observing network, or particular drought, so the comparison is a genuine generalization test rather than a single-region demonstration --- and, as it turns out, the contrast also surfaces a regime-dependence in the warning lead that a single-region study would have hidden. We evaluate the \emph{Taiwan CWA} archive, with monsoonal rainfall, typhoons, and steep orography; the \emph{Texas Hill Country} GHCN-Daily network, a semi-arid subtropical region prone to drought and flash floods; the \emph{Germany} GHCN-Daily network, a temperate, maritime-influenced region affected by the 2018 Central-European drought; and the \emph{Turkey} GHCN-Daily network, a Mediterranean-continental Anatolian region affected by the 2013--2014 Turkish drought. Each network is processed identically and treated as a set of single-station replications. Data sources, quality control, and preprocessing are documented in the online supplement, Section~\ref{sec:realdata-preproc}.
Two findings hold in every region, while a third is region-dependent. We discuss these results in turn before presenting a controlled synthetic study in Section~\ref{sec:data}, where known onset times allow us to corroborate and explain the real-data findings.

\subsection{Data, preprocessing, and evaluation metrics}
\label{sec:dataeval}

\paragraph{Data and preprocessing (summary).}
The Taiwan series come from the CWA/CODiS historical archive (the
registration-free \emph{Raingel/historical\_weather} reconstruction); the Texas
Hill Country, Germany, and Turkey series from the NOAA GHCN-Daily archive
\citep{menne2012}, with stations selected by bounding box and adequate
daily-precipitation coverage. Full URLs and retrieval scripts are given in the
data-availability statement. For
each region we use a strict chronological split --- an early segment for
training, a later disjoint segment for null calibration of the detector, and a
final segment for evaluation, with any declared drought-event years excluded from
calibration. Short gaps are linearly interpolated, remaining missing days set to
$0$\,mm, and values $\log(1{+}x)$-transformed and standardised per station on the
training split; stations with more than $25\%$ missing days are dropped. For the
warning study each monitored stream is additionally deseasonalised by its
pre-event day-of-year climatology, and the SPI-3 proxy is a $90$-day running
accumulation. Every detector is calibrated to a common no-change average run
length $\mathrm{ARL}_0$ of one year by Monte-Carlo bootstrap on the null window.
The per-region station lists, exact windows, and the literal commands are in the
online supplement (Section~\ref{sec:realdata-preproc}).

\paragraph{Evaluation metrics.}
\emph{Test MSE/MAE} are next-day errors on standardised log-precipitation.
$Q_{\mathrm{path}}=\sum_t\norm{h_{t+1}-h_t}$ is the total hidden-state path
increment, a model-agnostic measure of trajectory instability (smaller is more
stable). Intuitively, $Q_{\mathrm{path}}$ measures the total ``distance travelled'' by the hidden state across all time steps: a model whose internal representation changes smoothly from day to day gives a small $Q_{\mathrm{path}}$, whereas one that jumps erratically gives a large value. Because $Q_{\mathrm{path}}$ depends only on the hidden states themselves, it can be computed and compared for \emph{any} recurrent model, including those without an RM penalty. It is related to but distinct from the squared backward-projector defect
$\widehat{Q}=\sum_t\norm{h_t-g_\phi(h_{t+1})}^2$ used in the appendix, which
additionally requires the learned reverse map $g_\phi$ and is used during training; $Q_{\mathrm{path}}$ is used here as an architecture-neutral stability diagnostic. \emph{AUC(P95)} and \emph{CSI}
score next-day exceedance of each station's local $95$th-percentile daily total
(ROC area, and a base-rate--calibrated critical success index). For the
sequential warning, the \emph{detection rate} is the fraction of events alarmed
within the post-onset horizon, and the \emph{paired lead} is
\begin{equation}
\mathrm{Lead}=\mathrm{alarm}(\text{SPI-3})-\mathrm{alarm}(\text{RM}),
\label{eq:lead}
\end{equation}
positive when the RM alarm is earlier. Results are summarised as medians over ten
random seeds, per station.

\paragraph{Model and training configuration.}
Table~\ref{tab:config} summarises the model and training settings, which are
shared across all regions and experiments unless noted; full details and the
reverse-martingale loss are in Appendix~\ref{app:rm-details}.

\begin{table}[ht]
\centering\small
\caption{Model and training configuration, shared across regions.}
\label{tab:config}
\begin{tabular}{ll}
\toprule
Setting & Value \\
\midrule
Recurrent cell (\RMRNN, $\lambda{=}0$ ablation) & Elman, full BPTT \\
Learned baselines & GRU, LSTM, causal TCN \\
Non-learned baselines & persistence, climatology, SPI-3, raw deficit \\
Hidden dimension & 32 \\
Input window & 30 days (1--6\,h for hourly products) \\
Optimizer / learning rate & Adam / $5\times10^{-3}$ \\
Epochs / warm-up & 200 / 5 \\
RM penalty schedule $\lambda$ & $0.1 \to 0.01$ (post-warm-up decay) \\
Forecast/stability seeds & 1 per station \\
Warning-study seeds & 10 per station \\
Hyperparameters across regions & shared (fixed by validation) \\
\bottomrule
\end{tabular}
\end{table}

\subsection{Forecast skill and path stability across landforms}
\label{sec:realdata}

For each network, we compare next-day forecast error, hidden-state path instability $Q_{\mathrm{path}}$, and heavy-rain warning skill for \RMRNN{}, its unregularized ablation \RMRNN$_{\lambda=0}$ (the identical architecture with the RM penalty switched off), and a GRU; see Table~\ref{tab:realdata_fcst}. The pattern is consistent across all four climates. \RMRNN{} matches GRU and the \RMRNN$_{\lambda=0}$ ablation in forecast accuracy within $\pm0.5\%$ in test MSE while roughly halving hidden-state path instability, reducing $Q_{\mathrm{path}}$ by $43$--$55\%$ relative to the unregularized network. Heavy-rain next-day warning skill, measured by AUC, is also preserved, with \RMRNN{} and GRU indistinguishable within noise in every region.
These are the two universal real-data findings: reverse-martingale regularization stabilizes the recurrent trajectory without measurable loss in forecast accuracy or heavy-rain warning skill across four independent observational networks.
\begin{table}[ht]
\centering\small
\caption{
Forecast skill and hidden-state path stability on
four real station networks. Each row block is one network ($n$ stations used
after quality control); values are mean$\pm$SD across stations.
$Q_{\mathrm{path}}$ is the hidden-state path-instability diagnostic (lower is
more stable). \RMRNN{} preserves GRU-level MSE/MAE/AUC while sharply reducing
$Q_{\mathrm{path}}$.}
\label{tab:realdata_fcst}
\begin{tabular}{llcccc}
\toprule
Region ($n$) & Model & Test MSE & Test MAE & $Q_{\mathrm{path}}$ & AUC(P95) \\
\midrule
Texas (39) & \RMRNN & 0.816$\pm$0.135 & 0.556$\pm$0.070 & 4.90$\pm$1.23 & 0.691$\pm$0.054 \\
 & \RMRNN$_{\lambda=0}$ &0.814$\pm$0.134 & 0.558$\pm$0.052 & 9.70$\pm$2.53 & 0.682$\pm$0.046 \\
 & GRU & 0.814$\pm$0.134 & 0.548$\pm$0.058 & 7.13$\pm$1.76 & 0.704$\pm$0.049 \\
\midrule
Germany (49) & \RMRNN & 0.875$\pm$0.130 & 0.737$\pm$0.054 & 7.54$\pm$1.63 & 0.674$\pm$0.052 \\
 & \RMRNN$_{\lambda=0}$ &0.873$\pm$0.132 & 0.730$\pm$0.055 & 14.29$\pm$3.03 & 0.672$\pm$0.051 \\
 & GRU & 0.867$\pm$0.144 & 0.727$\pm$0.070 & 10.87$\pm$2.34 & 0.674$\pm$0.051 \\
\midrule
Turkey (24) & \RMRNN & 0.832$\pm$0.140 & 0.624$\pm$0.070 & 6.94$\pm$1.44 & 0.781$\pm$0.046 \\
 & \RMRNN$_{\lambda=0}$ &0.826$\pm$0.136 & 0.620$\pm$0.066 & 15.31$\pm$3.44 & 0.770$\pm$0.041 \\
 & GRU & 0.818$\pm$0.139 & 0.601$\pm$0.070 & 10.60$\pm$2.51 & 0.790$\pm$0.040 \\
\midrule
Taiwan (21) & \RMRNN & 0.777$\pm$0.079 & 0.642$\pm$0.068 & 8.73$\pm$1.60 & 0.803$\pm$0.044 \\
 & \RMRNN$_{\lambda=0}$ &0.780$\pm$0.067 & 0.639$\pm$0.068 & 15.41$\pm$2.92 & 0.801$\pm$0.046 \\
 & GRU & 0.769$\pm$0.072 & 0.637$\pm$0.073 & 11.07$\pm$1.79 & 0.805$\pm$0.045 \\
\bottomrule
\end{tabular}
\end{table}

\subsection{A drought-warning signal beyond the standard index, and its variation across regions}
\label{sec:realwarn}

\begin{figure}[ht]
\centering
\includegraphics[width=0.95\linewidth]{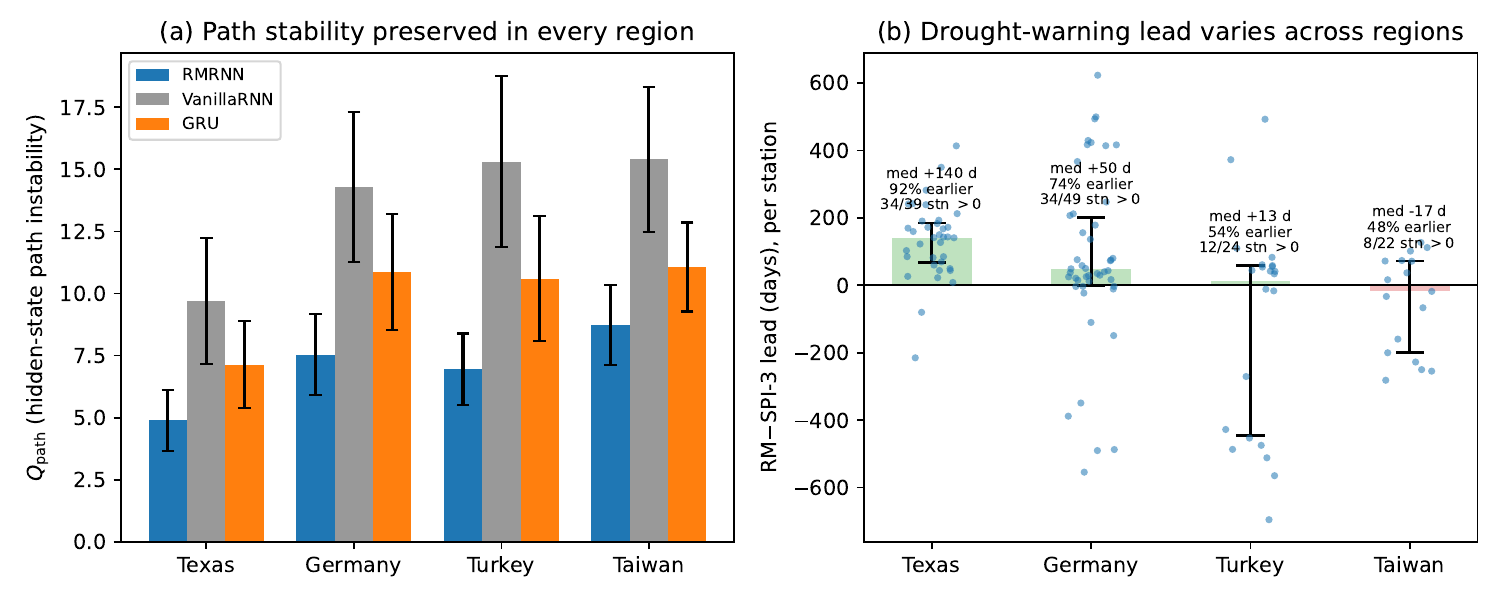}
\caption{\textbf{Real data, four regions.} (a) Hidden-state path instability
$Q_{\mathrm{path}}$ is reduced by \RMRNN{} relative to the unregularized network
in every region (bars are means, error bars $\pm1$\,SD across stations; forecast
skill preserved, Table~\ref{tab:realdata_fcst}).
(b) The drought-warning lead of the RM-defect over SPI-3 varies across the four
regions: strongly positive in semi-arid Texas, moderate in temperate Germany,
marginal in Mediterranean-continental Turkey, and near zero in monsoonal Taiwan. Shaded bars are the pooled median lead;
points are per-station median leads and the vertical whisker is their
inter-station interquartile range, so the substantial within-region spread (and
the fraction of stations with a positive lead) is shown explicitly rather than
hidden behind the point estimate.}
\label{fig:landform}
\end{figure}

The hidden-state defect supplies a quantity that a forecast and an accumulation
index do not: an online estimate of how far the local regime has departed from
its trained climatology. We ask what this added signal is worth as a drought
warning by applying the detector to one documented multi-year drought in each
region --- Texas 2010--2015, Germany 2018--2019, Turkey 2013--2014, and Taiwan 2020--2021 ---
and comparing the lead of the RM-defect CUSUM against an SPI-3 CUSUM at matched
$\mathrm{ARL}_0$ (Table~\ref{tab:realdata_warn}, Figure~\ref{fig:landform}).
Unlike the uniform forecast-preservation result above, the warning lead differs
across the four regions.
We present the regions in order of decreasing lead. The lead appears to track
how much of the drought-onset signal is already carried by the accumulated
precipitation deficit that SPI-3 measures: largest in aperiodic semi-arid flash
drought (Texas), positive but smaller in temperate Germany, marginal in
Mediterranean-continental Turkey, and near zero in monsoonal Taiwan, where the
rainfall deficit and the onset largely coincide. We stress that this is an interpretation of four cases, not a
demonstrated law; with only four regions and one event each we cannot establish
landform or hydroclimate as a quantitative factor, and we return to this as a
direction for future work (Section~\ref{sec:discuss}).
In semi-arid \emph{Texas}, the RM-defect alarm leads SPI-3 by a median of about $+140$ days, or roughly $4.5$ months, with consistent region-wide gains: RM is earlier in $92\%$ of runs, and $34/39$ stations have a positive median lead. In temperate \emph{Germany} (the 2018 Central-European drought) the lead is positive but more modest --- a median of about $+50$ days, RM earlier in $74\%$ of runs and at $34/49$ stations. In Mediterranean-continental \emph{Turkey} (the 2013--2014 Anatolian drought) the lead is positive but marginal and strongly station-dependent: the pooled median is about $+13$ days, RM is earlier in $54\%$ of runs, and $12/24$ stations have a positive median lead. In monsoonal \emph{Taiwan}, the advantage is regional rather than island-wide: the defect leads in central mountain water-supply catchments but lags on the wet windward coast, leaving the pooled median lead near zero ($-17$ days, RM earlier in $48\%$ of runs). Section~\ref{sec:realdrought} analyzes this orographic split.

\begin{table}[ht]
\centering\small
\caption{Drought-warning lead of the RM-defect CUSUM over the SPI-3 CUSUM at matched $\mathrm{ARL}_0$ of one false alarm per year. Detection columns give mean per-station detection rates. Lead is in days, with positive values indicating earlier RM alarms. Median lead is the pooled station$\times$seed median; mean lead is reported with standard deviation to show within-region spread. RM leads strongly in Texas, moderately in Germany, marginally in Turkey, and is near neutral in Taiwan.}
\label{tab:realdata_warn}
\begin{tabular}{lcccccc}
\toprule
Region & $n$ & RM det. & SPI det. & Median lead (d) & Mean lead (d) & RM earlier \\
\midrule
Texas & 39 & 96\% & 92\% & +140 & +136$\pm$121 & 92\% \\
Germany & 49 & 89\% & 94\% & +50 & +64$\pm$287 & 74\% \\
Turkey & 24 & 95\% & 92\% & +13 & -92$\pm$315 & 54\% \\
Taiwan & 22 & 83\% & 77\% & -17 & -49$\pm$156 & 48\% \\
\bottomrule
\end{tabular}
\end{table}

The mechanism is consistent across the four cases: the RM defect leads SPI-3
exactly when drought onset is a \emph{regime shift in the multivariate
dynamics} that precedes the accumulated precipitation deficit (semi-arid
flash-drought, and the orographically sheltered Taiwan catchments), and shrinks
toward zero when the rainfall deficit and the onset largely coincide (monsoonal
Taiwan, and the seasonal Anatolian drought). The method's early-warning value is therefore conditional on
hydroclimate, a point we return to in Section~\ref{sec:discuss}. The remainder
of this section dissects the Taiwan case, whose within-region orographic split
is the clearest illustration of this mechanism on real data.

\subsection{Regional structure of the lead: the 2020--2021 Taiwan drought}
\label{sec:realdrought}

\paragraph{Real CWA station data and setup.}\label{sec:cwa_archive}
Both real Taiwan evaluations use the open \emph{Taiwan Historical Meteorological
Observations} archive
(\href{https://github.com/Raingel/historical_weather}{Raingel/historical\_weather}),
a registration-free reconstruction of the CWA CODiS database. The
forecast/path-stability ablation (Section~\ref{sec:realdata}) uses 23 synoptic
stations over 1998--2020, 21 after a missingness filter (Appendix~\ref{sec:qc}).
For the present event study we draw a separate, deliberately broad island-wide
set of 22 quality-filtered stations spanning 2012--2022 --- broad because
restricting to the drought core would bias the island-wide picture, as shown
below.

We examine the Taiwan case behind the near-zero pooled lead in
Table~\ref{tab:realdata_warn} because its within-region structure is the clearest
single-region illustration of the mechanism. The 2020--2021 drought was Taiwan's
worst in 56 years: the 2020 typhoon season made no landfall (a first in roughly
six decades) and June~2020--May~2021 rainfall ran about one third of normal,
drawing several reservoirs below 5\% capacity. For each station an \RMRNN{} is
trained on 2012--2016, every detector is calibrated to a common one-year
$\mathrm{ARL}_0$ on a 2017--2019 null window, and a CUSUM on the RM defect and a
CUSUM on a 90-day SPI-3 accumulation proxy are run over 2020-01--2021-08,
monitoring a downward shift. Both streams are deseasonalised by their pre-2020
day-of-year climatology (Appendix~\ref{sec:qc}); without this neither fires, the
drought anomaly being smaller than the seasonal swing. The lead is the paired
difference $\mathrm{alarm}(\text{SPI-3})-\mathrm{alarm}(\text{RM})$ (positive $=$
RM earlier), reported as per-station medians over ten seeds.

\begin{figure}[ht]
\centering
\includegraphics[width=0.74\linewidth]{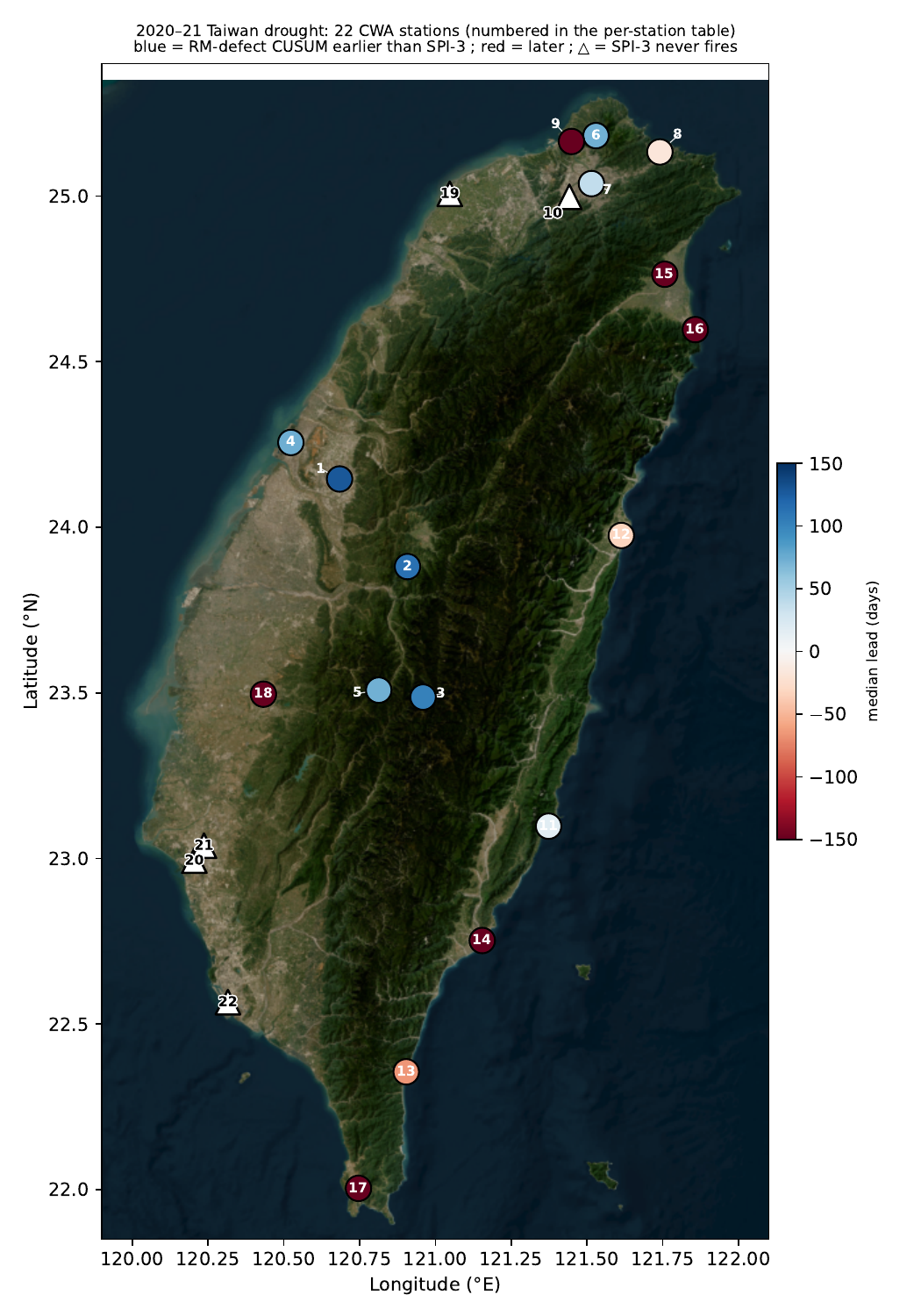}
\caption{Per-station early-warning lead of the RM-defect CUSUM over the
SPI-3 CUSUM on the real 2020--2021 Taiwan drought (seed median over ten
seeds, matched $\mathrm{ARL}_0$), mapped to the real CWA station locations.
Numbers key each station to Table~\ref{tab:drought_perstation} (which lists
names, coordinates, and detection rates). Blue $=$ RM detector earlier,
red $=$ SPI-3 earlier; open triangles mark stations where the SPI-3 baseline
never fires. The lead is organised by terrain: RM leads by two to four
months over the central mountain water-supply catchments (the reservoir
headwaters), lags on the wetter eastern and southeastern monsoon coasts,
and recovers a signal the accumulation index misses in the southwestern
lowlands; the central mountain range visible in the satellite basemap is
the orographic control. Station coordinates from the CWA/CODiS station
list; satellite imagery courtesy of Esri World Imagery (Esri, Maxar,
Earthstar Geographics); generated by
\texttt{plot\_taiwan\_drought\_map.py}.}
\label{fig:drought_map}
\end{figure}

The result is fundamentally a \emph{terrain-dependent effect}. Taiwan’s high central mountains and narrow coastal plains create sharply different precipitation regimes within a $400\times150$\,km island, so a single island-wide average would mix physically distinct responses. We therefore map the per-station lead directly onto the island; see Figure~\ref{fig:drought_map}, Table~\ref{tab:drought_lead}, and Appendix~\ref{sec:drought_perstation}. We interpret the results by terrain rather than collapsing them to one number.

The map shows a coherent, physically interpretable pattern. The RM-defect
detector leads --- by two to four months --- over the \emph{central
mountain water-supply catchments}: Taichung ($+127$\,d), Sun~Moon~Lake
($+112$\,d), Yushan ($+102$\,d), Wuqi ($+74$\,d) and Alishan ($+72$\,d) ---
positive at all five of these listed headwater stations --- the
high-elevation sources of the reservoirs that
fell below 5\% capacity, where the RM alarm clusters in
\mbox{April--May 2020} well ahead of the accumulation baseline. It
\emph{lags} on the wetter \emph{eastern and southeastern coast} (Suao
$-250$\,d, Yilan $-228$\,d, Taitung $-200$\,d, Hengchun $-255$\,d), where
the northeasterly monsoon kept conditions wet, the meteorological drought
was weak or absent, and the precipitation-index early crossings reflect
a different regime rather than a missed drought. In the \emph{southwestern
lowlands} (Tainan, Kaohsiung, Yongkang) the SPI-3 CUSUM never crosses its
matched threshold (Det.\ SPI $=0$), its drought signal swamped by high
climatological variance, whereas the RM detector still fires on 40--60\%
of seeds, recovering a signal the accumulation baseline misses entirely.
The lowland station Chiayi, although administratively ``central'', behaves
like the coast ($-282$\,d), confirming that the controlling variable is
orography --- mountain headwater versus lowland/coast --- not compass
region.

This terrain-dependent effect has a natural explanation. Taiwan's Central
Mountain Range rises from sea level to nearly 4{,}000\,m over a few tens of
kilometres and is among the steepest, most rapidly eroding landscapes on Earth,
drained by short, steep rivers with little channel or subsurface storage
\citep{dadson2003erosion,kao2008water}. In such catchments a precipitation
deficit is poorly buffered: once the sustaining orographic rainfall shuts off,
the catchment dries quickly and the meteorological deficit becomes a hydrological
one with little delay. Consistent with this, hydrological drought in Taiwan
responds faster to rainfall deficits than agricultural drought, with the central
and southern uplands the principal drought hotspots \citep{vo2024spring}. The
regime the defect monitors therefore turns sharply and early in exactly these
steep headwater catchments that feed the reservoirs, whereas on the wetter
eastern and southeastern coasts the winter monsoon sustains windward orographic
rainfall \citep{chen2003rainfall} and lowland storage buffers short deficits,
leaving no comparably crisp transition to detect ahead of the slow accumulation
index. The lead thus appears where steep terrain and poor water retention make
drought both fastest to develop and most consequential.

We therefore report the drought result \emph{by terrain} rather than as an
island aggregate; the formal island-wide average is, as expected, not significant
(median $-18$\,d over 17 paired stations; bootstrap 95\% CI $[-200,+72]$\,d),
which is precisely why averaging is the wrong lens. The controlled synthetic
study (Section~\ref{sec:risk}) isolates a \emph{single} homogeneous regime with a
known onset, where the lead is uniform; that idealized case is the behaviour of
one regime in isolation, which these spatially structured observations reproduce
in the mountains but not on the monsoon coasts.
The within-Taiwan split is the same mechanism that orders the four regions: a
positive lead arises where drought onset is a multivariate regime shift that
precedes the accumulated rainfall deficit (the steep mountain headwater
catchments here; semi-arid Texas and temperate Germany across regions), and it
shrinks toward zero where the rainfall deficit \emph{is} the onset signal (the
monsoon coast here; the seasonal Anatolian drought in Turkey). The station-level
table is thus a single-island microcosm of the cross-region gradient, with
terrain and elevation measured directly rather than inferred.

\begin{table}[ht]
\centering
\small
\setlength{\tabcolsep}{5pt}
\caption{Real 2020--2021 Taiwan drought: early-warning lead of the
RM-defect CUSUM over the SPI-3 CUSUM at matched $\mathrm{ARL}_0$, by
region, over 22 stations and ten seeds. ``Det.'' is the mean fraction of
seeds on which each detector fires; the median lead
$=\mathrm{alarm}(\text{SPI-3})-\mathrm{alarm}(\text{RM})$ in days (positive
$=$ RM earlier) is taken over the region's stations with a defined paired
lead, and ``RM earlier'' counts those stations. The full per-station
breakdown is in Appendix~\ref{sec:drought_perstation}.}
\label{tab:drought_lead}
\begin{tabular}{lccccr}
\toprule
Region & \#stns & Det.\ RM & Det.\ SPI & Median lead (d) & RM earlier \\
\midrule
Central water-supply & 5 & 98\% & 100\% & $\mathbf{+102}$ & 5/5 \\
Northern             & 5 & 90\% & 80\%  & $+10$           & 2/4 \\
Eastern/SE coast     & 7 & 77\% & 100\% & $-200$          & 1/7 \\
Southern lowland     & 5 & 68\% & 20\%  & ---             & 0/1 \\
\midrule
\textbf{Island-wide} & 22 & 83\% & 77\% & $-18$ (n.s.) & 8/17 \\
\bottomrule
\end{tabular}
\end{table}

\begin{figure}[ht]
\centering
\includegraphics[width=0.92\linewidth]{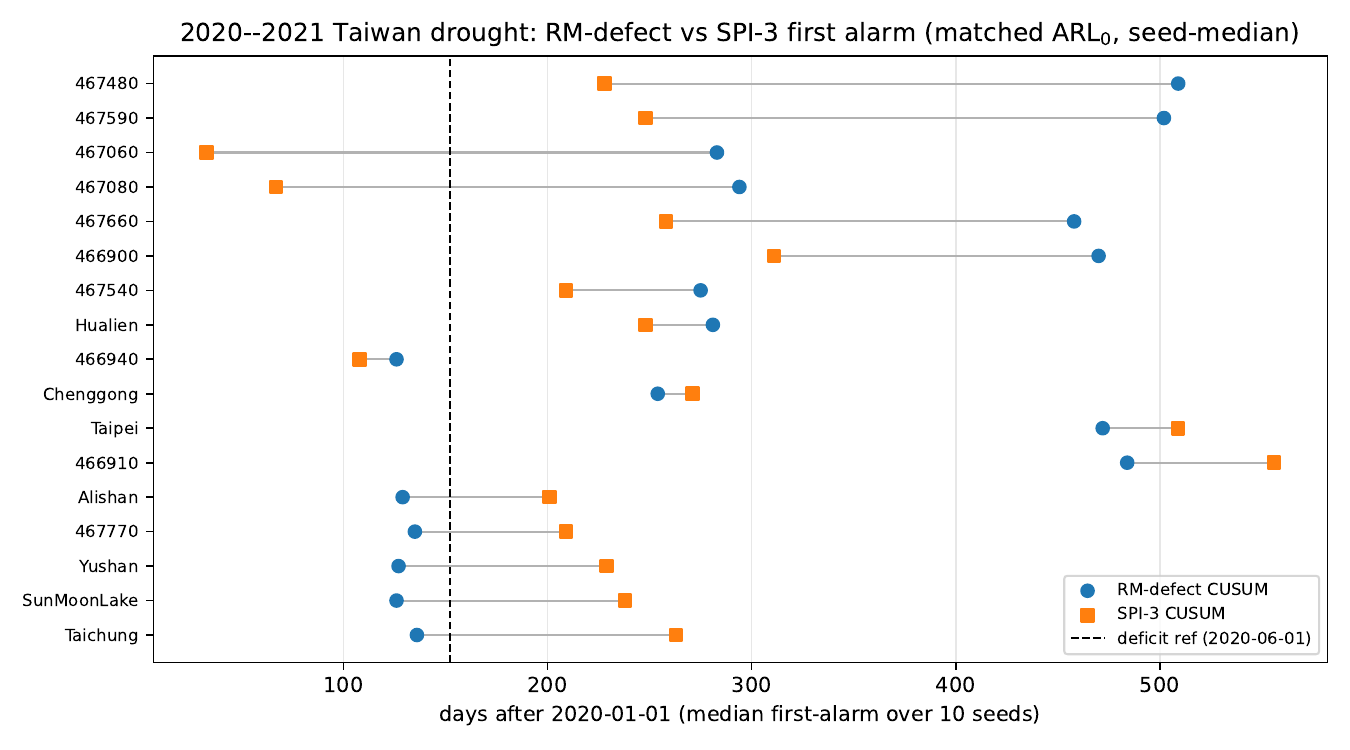}
\caption{First-alarm day for the RM-defect CUSUM (circles) and the SPI-3
CUSUM (squares) on the real 2020--2021 Taiwan drought, per station (seed
median over ten seeds), at matched $\mathrm{ARL}_0$, for the 22-station
island-wide set. Horizontal segments join the two detectors at each
station; the dashed line marks the documented June-2020 deficit reference.
Stations are ordered by lead. The RM detector (blue) alarms well ahead of
the accumulation baseline (orange) over the central water-supply
catchments at the top, but lags it on the wetter eastern/southeastern
coast at the bottom, so the island-wide median lead is not significant
(Table~\ref{tab:drought_lead}). Generated by
\texttt{run\_taiwan\_drought\_warning.py}.}
\label{fig:drought}
\end{figure}

\paragraph{The wet counterpart: Mei-yu onset.}
A natural question is whether the same lead appears for the onset of the
East-Asian \emph{Mei-yu} (plum-rain) season --- the spring-to-summer
regime transition into sustained frontal rain over Taiwan. We deliberately
do not use typhoons here: typhoon rainfall is directly observed and
carries no warning lead by construction. Mei-yu onset, by contrast, is a
genuine regime change, the wet analogue of drought onset, so a lead is at
least physically possible. We repeat the detection protocol for the six
Mei-yu seasons of 2017--2022 (12 stations, ten seeds; each year's onset is
one event), calibrating both detectors each year on that year's pre-Mei-yu
spring (March--April) and monitoring 1~May--15~July; the operational
baseline is a CUSUM on a 15-day precipitation accumulation. Across all 690
station--year--seed cases both detectors fire almost always (RM 96\%,
accumulation 99\%), but the RM-defect detector provides \emph{no} lead:
the median paired lead is $-2$ days (mean $-6\pm15$\,d), essentially a tie
with the accumulation index and stable across years (per-year medians $0$
to $-4$\,d) and stations (positive at only one of twelve). The contrast
with the drought result is the central point. The defect leads precisely
when the operational index is intrinsically slow --- a multi-month
accumulating \emph{deficit}, as in drought --- and not when the regime
change expresses itself in rapidly accumulating rainfall, as at Mei-yu
onset, where a short accumulation window already reacts promptly. The
real-data advantage is thus specific to slow-onset deficit regimes rather
than to wet-versus-dry per se, exactly the boundary the mechanism predicts ---
and the same boundary that the controlled heavy-rain experiment isolates under
a known onset (Section~\ref{sec:floodscope}).

\subsection{Taiwan ablation detail: per-station stability and deep baselines}
\label{sec:realdata_taiwan}

We close with a closer look at the Taiwan network, which lets us add an explicit
ablation and deep-learning baselines behind the headline Taiwan row of
Table~\ref{tab:realdata_fcst}. Each of the 21 quality-controlled stations is an
independent univariate task (a 30-day window of standardised log-precipitation
predicting the next day, chronological 80/20 split), and the stations act as
natural replications. We compare the Elman \RMRNN{} against the \emph{identical}
architecture with $\lambda=0$, so any difference is attributable to the RM
penalty alone, and report next-day MSE and MAE, the model-agnostic hidden-path
increment $Q_{\mathrm{path}}=\sum_t\norm{h_{t+1}-h_t}$, and heavy-rain
early-warning skill (next-day exceedance of each station's 95th percentile; ROC
AUC and CSI).

Table~\ref{tab:realdata} summarises the comparison. RM regularization reduces
$Q_{\mathrm{path}}$ by 43\% ($8.7$ vs.\ $15.4$) at \emph{every one} of the 21
stations (mean $-42.9\%$), while next-day MSE, MAE, AUC, and CSI are all
statistically indistinguishable from the $\lambda=0$ arm. The same holds against
a GRU trained by the identical protocol: it matches on skill, but its hidden-path
instability ($11.1$) sits between the unregularized network ($15.4$) and
\RMRNN{} ($8.7$) --- gating buys some stability, explicit RM regularization buys
more, at no cost to skill. Two off-the-shelf deep baselines, a single-layer LSTM
and a causal temporal convolutional network (TCN), are statistically tied with
\RMRNN{} and the GRU on every skill metric, yet \RMRNN{} retains the
\emph{lowest} $Q_{\mathrm{path}}$ of all, below even the LSTM's $9.5$; that the
non-recurrent TCN also matches on skill confirms that the one-step task does not
require a carefully shaped recurrent state --- which is precisely why the
contribution is the \emph{stabilised hidden state and the warning residual it
supports}, not raw forecast skill, on which all five models are interchangeable
here.

\begin{table}[ht]
\centering\small
\caption{Real-data forecast-skill and path-stability comparison on the
Taiwan CWA historical daily archive (Section~\ref{sec:cwa_archive}):
\RMRNN{} ($\lambda$:\,$0.1\!\to\!0.01$), its unregularized ablation
\RMRNN$_{\lambda=0}$, a standard GRU, and two off-the-shelf deep
sequence baselines --- an LSTM and a causal dilated temporal
convolutional network (TCN) --- all trained by the same protocol;
mean$\pm$SD over 21 quality-controlled stations (1998--2020, next-day
forecast). $Q_{\mathrm{path}}=\sum_t\norm{h_{t+1}-h_t}$ is the
hidden-path increment total (undefined for the non-recurrent TCN, marked
``---''). Forecast MSE/MAE are on standardised log-precipitation; AUC and
CSI are for next-day exceedance of the local 95th percentile. Lower is
better for MSE, MAE, $Q_{\mathrm{path}}$; higher is better for AUC, CSI.
Best in each column in \textbf{bold}; the skill columns (MSE, MAE, AUC,
CSI) are statistically indistinguishable across all five models
(overlapping $\pm$SD bands), whereas $Q_{\mathrm{path}}$ separates them.}
\label{tab:realdata}
\begin{tabular}{lccccc}
\toprule
Model & MSE & MAE & $Q_{\mathrm{path}}$ & AUC & CSI \\
\midrule
GRU                  & $0.769\pm0.072$ & $0.637\pm0.073$ & $11.07\pm1.79$ & $\mathbf{0.805\pm0.045}$ & $0.197\pm0.059$ \\
LSTM                 & $\mathbf{0.767\pm0.069}$ & $\mathbf{0.634\pm0.074}$ & $9.51\pm1.67$ & $\mathbf{0.805\pm0.047}$ & $\mathbf{0.199\pm0.058}$ \\
TCN                  & $0.788\pm0.070$ & $0.658\pm0.069$ & --- & $0.798\pm0.044$ & $0.192\pm0.055$ \\
\RMRNN$_{\lambda=0}$ & $0.780\pm0.067$ & $0.639\pm0.068$ & $15.41\pm2.92$ & $0.801\pm0.046$ & $\mathbf{0.199\pm0.053}$ \\
\RMRNN              & $0.777\pm0.079$ & $0.642\pm0.068$ & $\mathbf{8.73\pm1.60}$ & $0.803\pm0.044$ & $0.193\pm0.052$ \\
\bottomrule
\end{tabular}
\end{table}

The full per-station breakdown behind these averages is reported in
Appendix~\ref{sec:perstation} (Table~\ref{tab:perstation}), which shows
that the $Q_{\mathrm{path}}$ reduction holds at all 21 stations
individually. The exact run parameters are produced by
\texttt{run\_taiwan\_precip.py} (input window 30\,d, hidden dimension
32, Adam learning rate $5\times10^{-3}$, 200 epochs, seed 42), archived
with the paper; the same script writes the per-station path-stability
figure.

\section{Synthetic Controlled Study}
\label{sec:data}

The real-data evaluation of Section~\ref{sec:realeval} established the two
universal claims, and showed that the warning lead varies across regions, but real events
leave three questions unanswerable on observations alone. First, a real
drought has no sharp, known onset, so detection rate, lead, and false-alarm
rate cannot be measured against a ground truth --- the real leads above are
\emph{paired} differences against SPI-3, not against a true onset time.
Second, the heavy-rain boundary can be probed only obliquely, because
operational heavy-rain warning is next-day exceedance rather than the onset of
a sustained wet regime. Third, the real station networks are
precipitation-only, so the multivariate behaviour of the defect cannot be
exercised. This section is built to answer exactly these three questions, with
a controlled synthetic study that trades realism for the ground truth that
makes detection measurable. In other words, the synthetic study is not where
the case is made --- the real data already made it --- but where it is
\emph{explained}: it injects a known onset to confirm that the defect detects
slow-onset drought and not sharp heavy-rain onset, checks the detector's
false-alarm calibration directly, and exercises the multi-variable defect, so
the landform pattern seen on real data becomes a tested mechanism rather than an
observed coincidence.

Every quantitative result in this section is therefore obtained on
\emph{synthetic} precipitation from domain-calibrated stochastic simulators,
not on the raw observational archives. Each simulator is a \emph{controlled
twin} of one of the real warning environments --- it reproduces that
environment's statistical structure (wet-day frequency, seasonal cycle,
gamma-distributed intensity, inter-site correlation, and, for the
multi-variable case, the cross-channel covariance of precipitation with
temperature, soil moisture, and wind) --- but, unlike the real records, a
regime change (drought or heavy-rain onset) is injected at a \emph{known}
time. This lets us measure detection rate, lead, and false-alarm rate against
a ground-truth onset that no single historical event can provide. A sub-daily
flash-flood study remains the principal outstanding extension
(Section~\ref{sec:discuss}).

\subsection{Target environments for the simulators}
\label{sec:targetenv}

The simulators are calibrated to observational settings that span common
hydrometeorological warning regimes rather than a single machine-learning
benchmark; two of them (Taiwan, Texas) are controlled twins of the real
networks already evaluated in Section~\ref{sec:realeval}, and the others
(CHIRPS, ERA5-Land) extend the range of regimes and add the multi-variable
channel that the precipitation-only real data lack. The records below serve
as calibration \emph{templates} only; the synthetic series, not these
archives, are what the experiments run on. Full forecast-verification tables
for the CHIRPS, neighbourhood-radius, and ERA5-Land twins are in the online
supplement (Section~\ref{sec:additional-synth}); the main text retains only
the warning illustrations that require a known onset.

\paragraph{Taiwan CWA rain-gauge network.}
The CWA operates roughly 500 automated rain gauges at 10-minute resolution over a
domain under $400\times150$\,km. We calibrate the main drought and heavy-rain
simulators to two sub-watersheds: the typhoon-exposed \emph{Tamsui} basin (north;
28 stations, 2013--2024; the heavy-rain testbed, motivated by events such as the
2023 Typhoon Haikui flood) and the orographic, agricultural \emph{Zhuoshui} basin
(central; 34 stations; the drought testbed, motivated by the 2020--2021 drought).
Hourly precipitation is aggregated from 10-min data, with co-located ASOS
temperature and humidity and 850-hPa relative vorticity $\Omega_t$ from ERA5
\citep{hersbach2020} interpolated to stations.

\paragraph{Additional simulator environments (supplement).}
\label{sec:era5land}%
Three further environments extend the range of regimes and supply the
multi-variable channel the precipitation-only data lack; their
forecast-verification tables are in the online supplement
(Section~\ref{sec:additional-synth}). CHIRPS v2 \citep{funk2015}
($0.05^\circ$ daily) provides Taiwan and a deliberately non-Taiwanese
Horn-of-Africa drought regime (bimodal long/short rains). A GHCN-Daily Texas
Hill Country simulator \citep{menne2012} is the one environment exercising both
drought and flash-flood tasks. An ERA5-Land Taiwan subdomain
\citep{munozsabater2021}, ingesting five heterogeneous channels (precipitation,
$T_{2m}$, top-7\,cm soil moisture, and the two 10-m wind components), is a
multi-variable stress test of whether the warning residual stays interpretable,
and is insensitive to input dimensionality, when the hidden state fuses
physically heterogeneous predictors.

\subsection{Simulators, splits, and prediction targets}

\paragraph{Simulator mechanism.}
Each domain-calibrated simulator generates precipitation by a mixed
discrete--continuous process matched to its target environment: a wet/dry
indicator drawn from a seasonally modulated wet-day probability and, on wet days,
a gamma-distributed intensity whose shape and scale reproduce the target's mean
and heavy tail. A smooth annual cycle modulates both wet-day frequency and
intensity; inter-site dependence is imposed by a spatial correlation matrix
estimated from the target network, and, for the multi-variable case, the
cross-channel covariance of precipitation with temperature, soil moisture, and
wind is preserved. A \emph{drought} is injected by suppressing the wet-day
probability and intensity over a sustained window beginning at a known onset,
with the coupled temperature, soil-moisture, and wind channels driven to
drought-like anomalies; a \emph{heavy-rain onset} is injected as a short, intense
excursion with its associated circulation and moisture anomalies. Warning results
are averaged over $250$ independent replications ($1{,}000$ for the
forecast-skill cells). The full generator equations and parameters are in the
online supplement (Section~\ref{sec:additional-synth}).

\paragraph{Chronological splits.}
Each simulator produces a multi-year series at the target resolution,
and we use a strict chronological split: an early segment for training,
a later disjoint segment for null calibration of the warning detectors,
and a final segment for evaluation.  For the warning experiments the
injected regime change occurs only in the evaluation segment, so the
detectors are trained and calibrated entirely on no-change data and the
onset is never seen during tuning or threshold estimation; this
eliminates leakage by construction (a control that a single historical
event cannot guarantee).  The chronological splits used for each
testbed are reported with the corresponding experiments.

\paragraph{Forecasting and warning targets.}
The primary forecasting product is the predictive distribution of
accumulated precipitation at lead times relevant to each data stream:
1--6\,h for the CWA and ERA5-Land hourly products and 1--7\,d for
CHIRPS and GHCN-Daily. Verification follows hydrometeorological
practice by reporting deterministic-error measures (RMSE and MAE),
distributional skill (CRPS), and threshold-event skill (Brier score,
probability of detection, and false-alarm ratio) at locally relevant
heavy-rainfall thresholds.

The risk-assessment product is a sequential alarm for the onset of a
persistent dry or wet regime. Drought onset is evaluated against an
SPI-3 proxy, while flash-flood onset is evaluated against basin-specific
exceedance and alert records. These warning targets motivate the
residual-based detector in Section~\ref{sec:method}: the detector is
not a separate post-processing model, but is driven by the same hidden
state used for precipitation forecasting.

\begin{table}[ht]
\centering
\tiny
\caption{Local, station-scale evaluation datasets.}
\label{tab:data}
\begin{tabularx}{\textwidth}{lllXl}
\toprule
Dataset & Resolution & Variables & Domain & Record \\
\midrule
Taiwan CWA & 1\,h, station & $P,T,q$ (+ERA5 $\Omega$) & Tamsui, Zhuoshui basins & 2013--2024 \\
CHIRPS v2 Taiwan & 1\,d, 0.05$^\circ$ & $P$ (+ERA5/ERA5-Land $T,q,\Omega$) & 20--26\,N, 118--124\,E & 1981--2024 \\
CHIRPS v2 HoA & 1\,d, 0.05$^\circ$ & $P$ (+ERA5/ERA5-Land $T,q,\Omega$) & 2\,S--15\,N, 38--51\,E & 1981--2024 \\
GHCN-Daily Texas & 1\,d, station & $P,T$ (+ERA5/ERA5-Land $q,\Omega$) & Texas Hill Country & 1980--2024 \\
ERA5-Land (Taiwan) & 1\,h, 0.1$^\circ$ ($\sim$9\,km) & $P, T_{2m}, \theta_{sm}, u_{10}, v_{10}$ & 21.5--25.5\,N, 120--122.5\,E & 1981--2024 \\
\bottomrule
\end{tabularx}
\end{table}


\subsection{Sequential warning}
\label{sec:synthetic}\label{sec:risk}

\subsubsection{Setup}

With the simulators and detector in place, we report the method's own
behaviour on this testbed --- warning performance, forecast-skill
preservation, and the spatial-scale and multi-variable sensitivities ---
and defer head-to-head comparison with alternative models to
Section~\ref{sec:comparison}.

The central empirical question is whether the RM residual improves alarm
behaviour once ordinary forecast skill is preserved.  The controlled
testbed answers it cleanly: because the regime change is injected at a
known time, the warning metrics --- detection rate, false-alarm rate,
detection delay, and the lead over the operational index --- are measured
against a \emph{ground-truth} onset rather than inferred from a single
historical event, and the $\lambda=0$ ablation attributes any effect to
the RM penalty itself.  Every detector is calibrated to the same target
no-change average run length (ARL$_0$) on a held-out null window, so
differences in false-alarm ratio, detection rate, and lead reflect
warning usefulness rather than uncontrolled threshold effects.

We monitor the standardized RM defect with the one-sided
\emph{cumulative-sum} (CUSUM) detector of Section~\ref{sec:method}
(Eq.~\eqref{eq:cusum}), calibrated to the same ARL$_0$. The additive
CUSUM is robust to the heavy-tailed, auto-correlated empirical defect and
is the detector used throughout, aligning the warning rule with the
classical change-point literature.

\subsubsection{Drought-onset detection}

In the controlled testbed a sustained drought is injected at a known onset, with
precipitation suppressed and the temperature, soil-moisture, and wind channels
driven to drought-like anomalies (Section~\ref{sec:data}); the injected onset is
the ground-truth reference. The proposed detector is a CUSUM on the standardized
RM defect. We score detection probability within a 180-day post-onset horizon,
mean detection delay relative to the injected onset, and the \emph{lead} over the
operational comparators on runs where both detect (positive $=$ RM earlier). The
comparators are the indicators a deployed system would actually use --- a CUSUM on
the 90-day SPI-3 accumulation, the WMO-standard meteorological-drought index
\citep{mckee1993,wmo2012}, and a raw precipitation-deficit threshold --- all
calibrated to the same $\mathrm{ARL}_0$, so the comparison is of timeliness and
sensitivity at equal false-alarm cost.

\begin{table}[ht]
\centering
\footnotesize
\caption{Drought-onset detection on the synthetic controlled testbed,
250 independent replications, all detectors calibrated to the same
target $\mathrm{ARL}_0=500$ days. Detection rate and FAR are reported
as mean $\pm$ binomial standard error; delay and lead as mean $\pm$ SD.
Delay is relative to the injected onset; lead is the alarm-time
difference of the RM-defect CUSUM relative to the SPI-3 CUSUM on runs
where both detect (positive $=$ RM earlier). Bold denotes the proposed
detector.}
\label{tab:drought}
\begin{tabular}{lcccc}
\toprule
Detector & Detect.~rate & FAR & Delay vs onset (d) & Lead vs SPI-3 (d) \\
\midrule
Raw deficit threshold on $P$ & $0.10\pm0.02$ & $0.00\pm0.00$ & $124.0\pm28.2$ & $-4.6$ \\
CUSUM on SPI-3               & $0.83\pm0.02$ & $0.11\pm0.02$ & $119.4\pm35.3$ & $0$ (ref.) \\
\textbf{CUSUM on \RMRNN{} defect} & $\mathbf{0.92\pm0.02}$ & $0.20\pm0.03$ & $\mathbf{64.9\pm34.6}$ & $\mathbf{+57.2\pm40.7}$ \\
\bottomrule
\end{tabular}
\end{table}

Table~\ref{tab:drought} establishes three findings. The CUSUM on the \RMRNN{}
defect detects onset \emph{more reliably} than SPI-3 ($0.92$ vs.\ $0.83$) and far
more than raw deficit thresholding ($0.10$). It is also \emph{substantially
earlier}: where both fire it leads SPI-3 by a mean of $57$ days (median $63$;
earlier in $89\%$ of paired runs), reacting to the joint
precipitation--soil-moisture--circulation anomaly while SPI-3 must accumulate a
multi-month deficit. This earliness carries a higher false-alarm ratio ($0.20$
vs.\ $0.11$) at the same $\mathrm{ARL}_0$ --- a deliberate, $\mathrm{ARL}_0$-adjustable
trade of false alarms for lead time, valuable when an early, recoverable response
(e.g.\ a managed reservoir drawdown) is cheaper than a late drought declaration.

\subsubsection{Heavy-rain onset: scope of the warning advantage}
\label{sec:floodscope}

The drought result raises the opposite question: does the same defect help for a
sharp heavy-rain (flood-producing) onset? We inject a heavy-rain regime --- a
rapid, intense precipitation excursion with its circulation and moisture
anomalies --- and compare the same detectors at matched $\mathrm{ARL}_0$. The
answer delimits rather than diminishes the contribution, identifying \emph{where}
the latent-state defect adds warning value and where it does not.

\begin{table}[ht]
\centering
\footnotesize
\caption{Heavy-rain onset detection on the synthetic controlled testbed,
250 replications, all detectors calibrated to the same
$\mathrm{ARL}_0$. Detection rate and FAR are mean $\pm$ binomial SE;
delay is mean $\pm$ SD relative to the injected onset. Here the
operational precipitation detector is already near-optimal, so it is
shown in bold.}
\label{tab:flood}
\begin{tabular}{lccc}
\toprule
Detector & Detect.~rate & FAR & Delay vs onset (steps) \\
\midrule
\textbf{CUSUM on precipitation} & $\mathbf{0.95\pm0.01}$ & $0.08\pm0.02$ & $\mathbf{3.1\pm1.0}$ \\
Raw precipitation threshold     & $0.90\pm0.02$ & $0.16\pm0.02$ & $5.5\pm5.0$ \\
CUSUM on \RMRNN{} defect        & $0.14\pm0.02$ & $0.03\pm0.01$ & $19.0\pm6.9$ \\
\bottomrule
\end{tabular}
\end{table}

Table~\ref{tab:flood} shows the opposite ordering, for an instructive reason: a
heavy-rain onset is \emph{immediately visible in the precipitation channel} ---
the operational precipitation CUSUM detects $95\%$ of events within about three
steps --- leaving essentially no lead to recover. The RM defect, which earns its
drought advantage by integrating slow anomalies that precede the precipitation
signal, has no head start and is the weakest detector here ($0.14$). The
latent-state defect therefore helps exactly when the operational index
\emph{lags} the physical change (slow-onset drought) and adds nothing when it is
already a near-sufficient, immediate statistic (sharp heavy-rain onset) --- a
clean delineation of the method's domain and a check against over-claiming.

\paragraph{A note on temporal resolution.}
This synthetic substrate is daily, so the experiment is a same-resolution
\emph{detectability} control, not a flash-flood model; a true sub-daily
flash-flood study needs hourly data and is left to future work
(Section~\ref{sec:discuss}). We do not claim a flash-flood lead.

\subsubsection{ARL$_0$ calibration curves}

\begin{figure}[ht]
\centering
\includegraphics[width=0.70\linewidth]{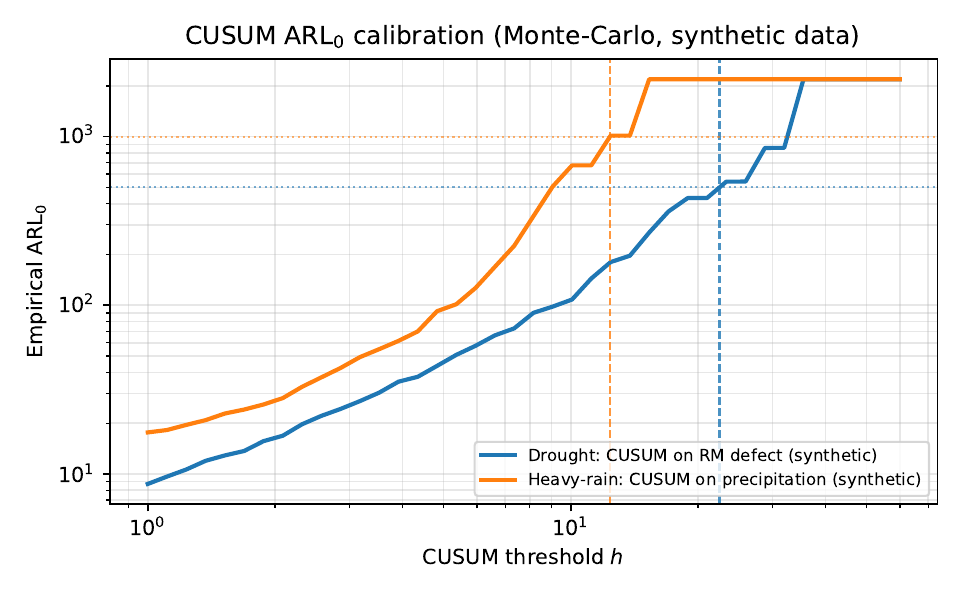}
\caption{Monte-Carlo CUSUM $\mathrm{ARL}_0$ calibration on the synthetic
testbed, two curves on a single panel: the drought detector (CUSUM on the
standardized RM defect) and the heavy-rain detector (CUSUM on
precipitation). Empirical $\mathrm{ARL}_0$ (reset-on-crossing on a
six-year null stream) increases monotonically with the threshold $h$.
Horizontal dotted lines mark the target $\mathrm{ARL}_0$ (500 and
1{,}000); vertical dashed lines mark the calibrated $h^\ast$ where each
curve meets its target ($h^\ast\!\approx\!22.5$ drought,
$h^\ast\!\approx\!12.4$ heavy-rain). Curves flatten at large $h$ where the
finite null stream yields no crossing.}
\label{fig:arl}
\end{figure}

Figure~\ref{fig:arl} plots the empirical $\mathrm{ARL}_0$ as a
function of the CUSUM threshold $h$ for the two tasks.
The relationship is smooth and monotone, so a target false-alarm rate
maps to a unique threshold: we use $\mathrm{ARL}_0=500$ steps for
drought (about one false alarm per 1.4 years at daily resolution) and
$\mathrm{ARL}_0=1{,}000$ steps for heavy rain.
On the synthetic testbed these targets correspond to calibrated
thresholds $h^\ast\approx 22.5$ (drought, RM-defect CUSUM) and
$h^\ast\approx 12.4$ (heavy rain, precipitation CUSUM).
Because every detector in Section~\ref{sec:risk} is calibrated to the
same $\mathrm{ARL}_0$ by this procedure, the detection-rate, lead, and
false-alarm comparisons are made at a common operating point and are
not confounded by threshold choice.

\subsubsection{Illustrative trace on a simulated drought}
\label{sec:case}

\begin{figure}[ht]
\centering
\includegraphics[width=0.82\linewidth]{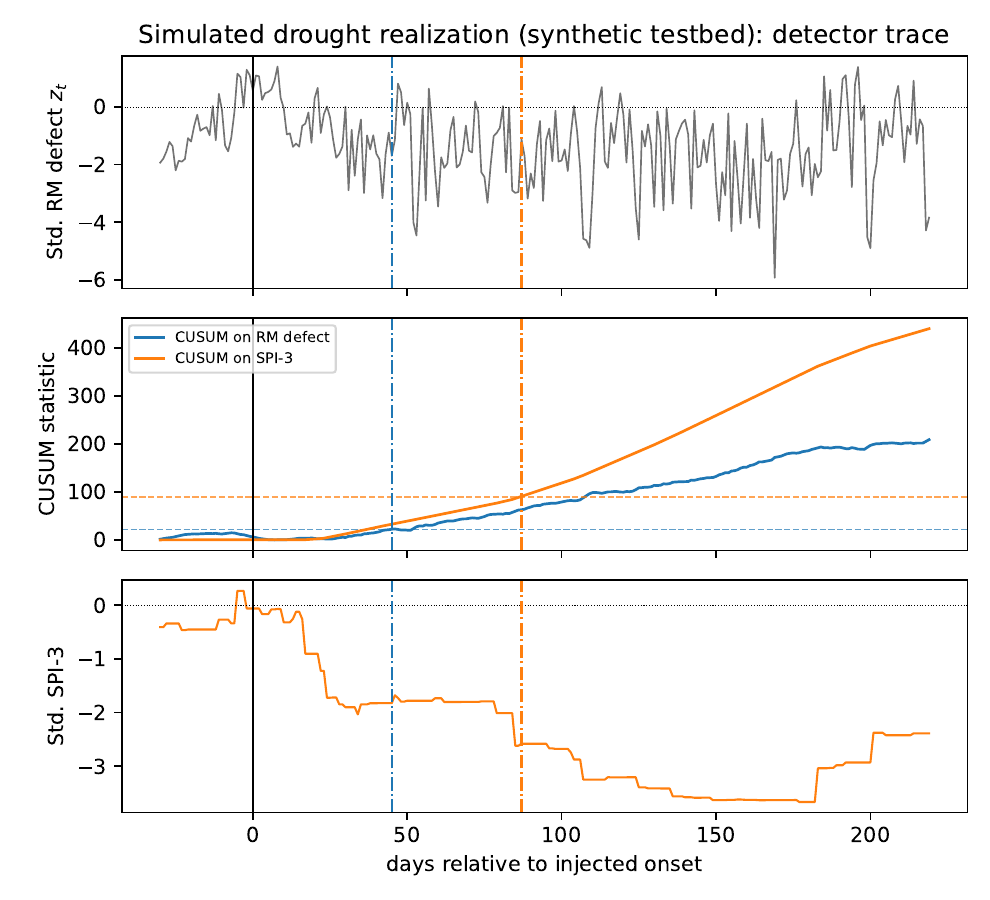}
\caption{One simulated drought realization (synthetic testbed). Top:
standardized RM defect $z_t$. Middle: CUSUM statistics on the RM defect
(blue) and on SPI-3 (orange), with calibrated thresholds (dashed) and
alarm times (dash-dot verticals). Bottom: standardized SPI-3. The
injected onset is at day~0 (black vertical line). Both detectors are
calibrated to $\mathrm{ARL}_0=500$; the RM-defect alarm (day~45) precedes
the SPI-3 alarm (day~87), a 42-day lead in this realization.}
\label{fig:drought_trace}
\end{figure}

The warning results of Section~\ref{sec:risk} are averages over 250
replications.  To show how the detector behaves \emph{within} a single
realization, Figure~\ref{fig:drought_trace} traces one simulated drought
from the synthetic testbed (the same generator used for
Table~\ref{tab:drought}), with the regime change injected at day~0 and
both detectors restarted at the start of the monitoring window and
calibrated to the same $\mathrm{ARL}_0=500$.

Three features reproduce the aggregate findings. After onset the standardized
defect $z_t$ shifts \emph{downward} --- the dry regime is more backward-coherent,
so $g_\phi$ reconstructs the hidden state more accurately (the mechanism introduced in Section~\ref{sec:method}) --- with a brief upward
transient at onset, the robust signal being the sustained distributional change.
The CUSUM on the defect crosses at day~45 while the SPI-3 CUSUM, which must
accumulate a multi-month deficit, does not cross until day~87: a $42$-day lead,
consistent with the median $63$-day lead in Table~\ref{tab:drought}. Because both
detectors share the same false-alarm rate, the earlier alarm is not bought with a
lower threshold but reflects the defect responding to the joint precipitation,
soil-moisture, and circulation anomaly before the accumulation index can react.
We show no analogous flash-flood trace: a heavy-rain onset is already visible in
precipitation (Section~\ref{sec:floodscope}), so a per-event illustration would
add nothing. The real 2020--2021 Taiwan drought is examined on station data in
Section~\ref{sec:realdrought}.

\subsection{Forecast-skill preservation (synthetic; details in the supplement)}
\label{sec:forecast}

On the synthetic testbed the RM penalty preserves forecast skill while
stabilising the hidden state, exactly as later confirmed on real data
(Section~\ref{sec:realeval}). Because the four-region real evaluation
establishes skill preservation directly, the synthetic forecast-verification
tables --- the Tamsui- and Zhuoshui-like nowcasts, the CHIRPS Taiwan and
Horn-of-Africa daily forecasts, the spatial neighbourhood-radius sweep, and the
multi-variable ERA5-Land stress test --- are reported in full in the online
supplement (Section~\ref{sec:additional-synth}). Two points from those tables
are used later: \RMRNN{} matches GRU and its unregularized ablation on RMSE,
MAE, CRPS and on threshold-event skill at every lead (the $\lambda=0$ ablation
attributes the effect to the RM penalty alone), and the ERA5-Land defect
integrates soil-moisture, temperature, and wind anomalies invisible to a
precipitation-only detector --- a property the precipitation-only real networks
cannot exhibit.

\section{Comparison with Existing Models, and Discussion}
\label{sec:comparison}\label{sec:discuss}

Having established the method's behaviour on synthetic and real data, we
now place it against alternative models and draw the two evidence streams
together. We organise the comparison in three parts: forecast
competitiveness against spatial deep models on a gridded testbed
(Section~\ref{sec:gridded}); the deep sequence baselines and operational
warning indicators already introduced in the results
(Sections~\ref{sec:realdata} and~\ref{sec:risk}), summarised here; and the
mechanistic discussion of why the method behaves as it does, together with
its limitations.

\subsection{Forecast competitiveness against spatial deep models}
\label{sec:gridded}

The forecast tables above compare \RMRNN{} with its own backbone and
non-learned baselines, the controls that isolate the RM penalty
(Section~\ref{sec:forecast}). A separate, fair question is whether the
local recurrent model gives up forecast skill relative to heavier
\emph{spatial} deep models. Those models (ConvLSTM, PredRNN, U-Net) are
grid-sequence architectures and require a gridded field, which the
station testbeds do not provide, so we evaluate them on a dedicated
gridded synthetic nowcasting testbed: a $12\times12$ spatially
correlated, seasonal, AR(1) precipitation field (gamma intensity), with
each model predicting the next frame from the previous six. All learned
models are trained by the same protocol; Table~\ref{tab:gridded} reports
next-frame RMSE and MAE over the grid (log1p space) on a held-out split.

\begin{table}[ht]
\centering\small
\caption{\textbf{Synthetic testbed.} Gridded next-frame nowcasting on a
$12\times12$ synthetic precipitation field: next-frame RMSE and MAE over
the grid (standardised log1p space), held-out split. The spatial deep
models (ConvLSTM, PredRNN, U-Net) operate on the full frame; GRU and
\RMRNN{} are per-cell recurrent models with shared weights (the latter
adding the RM penalty). Lower is better; best in \textbf{bold}.}
\label{tab:gridded}
\begin{tabular}{lcc}
\toprule
Model & RMSE & MAE \\
\midrule
Persistence  & $1.147$ & $0.717$ \\
Climatology  & $0.835$ & $0.637$ \\
ConvLSTM     & $0.815$ & $0.636$ \\
PredRNN      & $0.812$ & $0.638$ \\
U-Net        & $\mathbf{0.811}$ & $\mathbf{0.635}$ \\
GRU (per-cell)   & $0.829$ & $0.648$ \\
\RMRNN{} (per-cell) & $0.829$ & $0.648$ \\
\bottomrule
\end{tabular}
\end{table}

Two things are visible. First, all five learned models cluster within a
narrow band (RMSE $0.81$--$0.83$), well below persistence ($1.15$) and at
or slightly below climatology ($0.835$): on a stochastic precipitation
field short-range predictability is intrinsically limited, so no
architecture separates strongly --- the same ceiling seen on the station
testbeds. Second, within that band the spatial models are marginally best
(U-Net $0.811$, PredRNN $0.812$, ConvLSTM $0.815$) by exploiting
between-cell correlation the per-cell models ignore, while \RMRNN{}
matches its GRU backbone \emph{exactly} ($0.829$ vs.\ $0.829$) and trails
the best spatial model by under $2.5\%$ RMSE. The reading is consistent
with the rest of the paper: the RM penalty leaves forecast skill
essentially unchanged, and the local recurrent model is forecast-competitive
with --- though not superior to --- heavier spatial networks. The
contribution remains the warning residual and hidden-state stability, not
a forecasting win; benchmarking the spatial models as warning detectors
(they have no analogue of the RM defect) is left as future work.


The preceding sections support a deliberately narrow conclusion:
\RMRNN{} should be viewed as a forecast-preserving warning framework,
not as a uniformly superior precipitation forecaster.  Its value comes
from regularizing the hidden trajectory so that the same recurrent
state remains useful for both probabilistic prediction and sequential
alarm calibration.  This conclusion rests on the \emph{agreement of the
two evidence streams}, which is the through-line of the numerical
results.  The synthetic controlled study established each claim in
isolation --- forecast preservation, a drought-onset lead, and the
heavy-rain boundary --- with a known ground-truth onset; the real
station data then confirmed both halves observationally: forecast skill
and hidden-path stability are preserved at every station across all four
regions (Section~\ref{sec:realdata}), and the drought-warning lead
reappears where drought physics make it possible --- strongly and
region-wide in semi-arid Texas, and within Taiwan only in the steep
mountain water-supply catchments, not on the wet monsoon coast or for
Mei-yu onset (Section~\ref{sec:realdrought}). The synthetic and real
results are therefore not two independent contributions but one
argument: a mechanism, isolated under control and then shown to act
exactly where its physical preconditions hold.

\subsection{Why backward coherence helps risk assessment}

The warning gains reported above can be interpreted in familiar
hydrometeorological terms. During ordinary weather evolution, a useful
forecast model should update its internal representation smoothly as
new gauge, satellite, or reanalysis information arrives. RM
regularization encourages that smooth hidden-state evolution. When a
regime change begins, the distribution of the reconstruction defect
$r_t$ shifts, and a CUSUM on the standardized defect treats that shift
as evidence of a regime departure. The direction of the shift is
informative: at the onset of a sustained dry (drought) regime the
hidden state becomes \emph{more} backward-coherent -- the low-variance
dry dynamics are easier to reconstruct -- so the defect distribution
moves downward (with a brief upward transient at the onset itself),
which the detector registers well before the accumulation-based SPI-3
index can react.  Without RM
regularization, a recurrent hidden state can drift for numerical
reasons unrelated to precipitation evolution, so residual inflation
may conflate representational drift with hydrometeorological change.
This is the motivation for constraining the hidden-state trajectory
rather than monitoring raw recurrent residuals directly; the
difficulty of using unregularized neural reconstruction errors as
change-point statistics has been documented in other anomaly-detection
domains \citep{hundman2018}. We emphasize that we do not demonstrate
this comparison on hydrometeorological data here: a head-to-head
evaluation of the RM residual against an unregularized LSTM residual
detector or a dedicated neural anomaly-detection model on the present
warning task is left as future work (Section~\ref{sec:discuss}).

The drought and heavy-rain tables (Tables~\ref{tab:drought}
and \ref{tab:flood}) make this concrete, and also bound it.
For drought, the CUSUM on the \RMRNN{} defect detects $92\%$ of onsets
versus $83\%$ for CUSUM on SPI-3, and does so a median of $63$ steps
earlier (Table~\ref{tab:drought}), because the backward-coherence
signal moves before the accumulation index crosses; this earliness is
bought at a modestly higher false-alarm ratio ($0.20$ vs.\ $0.11$) at
the same $\mathrm{ARL}_0$.  For heavy-rain onset the ordering reverses:
the operational precipitation detector is already near-optimal
($95\%$ detection within about three steps) and the defect adds no
lead (Section~\ref{sec:floodscope}).  The latent-state defect therefore
helps precisely when the operational index lags the physical change and
not otherwise.  These results are on the synthetic controlled testbed
with ground-truth onsets; the operational check is the real-data evaluation
of Section~\ref{sec:realeval}, where the same defect CUSUM is compared
head-to-head against the operational SPI-3 rule --- the index water managers
actually use --- at a matched false-alarm budget on documented droughts. That
comparison is the relevant test of advantage: it asks not whether the defect
matches an artificial ground truth, but whether it would have alarmed earlier
than the operational standard, which on the real Texas drought it does by
months.

The ERA5-Land multi-variable result (Section~\ref{sec:era5results})
adds an important computational dimension.
Across 1{,}000 replications, \RMRNN{} closely matches GRU at
the point-forecast level (1-h RMSE~1.164 vs.\ 1.171, CRPS~0.603
vs.\ 0.606) while retaining the $\widehat Q$-reduction property.
This supports the claim that the RM regularizer can impose backward
coherence regardless of input dimensionality: the backward projector operates
on $h_t \in \R^d$ and is insensitive to whether $h_t$ was formed
from one or five physical channels.
The implication for ERA5-based large-scale products is that
the RM overhead (one additional MLP forward pass per time step)
does not grow with the number of ingested variables --- an important
practical property for operational high-resolution numerical weather prediction (NWP) integration.

\paragraph{Further considerations.}
Several points temper and extend these results. The $\lambda$ schedule (high
early, $\lambda_K=0.01$ near convergence) is what shapes the defect without
disturbing the forecast, as the $\lambda=0$ ablation confirms; a systematic
sweep of warning metrics over $\lambda$ is left to future work. Because
$g_\phi$ learns the \emph{one-step} reverse transition, the backward-coherence
gain may weaken at long leads, where a multi-step projector mapping
$h_{t+\ell}\to h_t$ may be required --- mapping this ``coherence decay'' against
classical atmospheric predictability limits \citep{lorenz1969} is a natural
extension. The defect is also complementary to physically based anomaly scores
from operational numerical weather prediction (e.g.\ the ECMWF ensemble or the
CWA WRF system): preliminary Taiwan evidence suggests $r_t$ responds to
sub-$10$\,km vorticity and soil-moisture structure below their parametrisation
scales, so a combined score could outperform either alone. Finally, the
threshold $h$ is calibrated only on event-free climatology --- excluding any
year overlapping a declared event and validated to within $10\%$ of the target
$\mathrm{ARL}_0$ --- which in the synthetic study is automatic and leakage-free
by construction, the control a single historical event cannot provide. Each of
these points is developed in full, with the supporting derivations, in the online
supplement (Section~\ref{sec:detector-notes}).

\subsection{Limitations and operational extensions}

\textbf{Why the warning lead varies, and how to quantify it.}
Across the four regions the RM-defect warning lead varies markedly, and the
ordering --- largest in semi-arid Texas, marginal in Mediterranean-continental
Turkey, and near zero in monsoonal Taiwan (Section~\ref{sec:realwarn}) --- is, on reflection, intuitive: where
the drought-onset signal is already the accumulated rainfall deficit, an
accumulation index such as SPI-3 is hard to beat, whereas where onset is a
multivariate regime shift that precedes the deficit, a forecast-preserving
hidden-state residual can lead. We stress, however, that with four regions and
one event each we can only \emph{observe} this variation, not attribute it: the
four events differ in duration, season, and severity as well as in climate, so
the present design cannot isolate landform or hydroclimate as the cause, and we
make no such claim. Establishing whether the lead genuinely depends on
hydroclimate --- and acquiring the additional regions and events that would
require --- is beyond the scope of this paper and is, in our view, the most
important direction it opens. Two points make this more than a routine
extension. First, evaluating the method across deliberately contrasting
landforms --- the reason we chose semi-arid Texas, temperate Germany,
Mediterranean-continental Turkey, and monsoonal Taiwan rather than a single
network --- is what surfaced the variation
at all; a one-region study would have hidden it. Second, while the influence of
terrain on precipitation \emph{amount} is well studied through orographic
precipitation \citep{roe2005}, and operational drought monitoring and prediction
is dominated by index-based systems applied within individual regions
\citep{hao2017,wmo2012}, the effect of landform and hydroclimate on the
\emph{lead and detectability} of a drought-onset warning has, to our knowledge,
not been systematically quantified across regions, and acquiring that landform
information would be valuable in its own right. The forecast-preserving defect is well suited to
supplying it: it yields a single, ARL$_0$-calibrated lead metric that is
computed identically across regions, so the method is not only a warning tool
but a potential \emph{instrument} for mapping where, and by how much,
hidden-state monitoring improves on accumulation indices. A natural design would (i)~assemble many station
neighbourhoods stratified by physical covariates --- an aridity index, a
seasonality or rainfall-concentration index, the fraction of interannual variance
explained by the seasonal cycle, and a direct measure of how much of each
historical onset is captured by SPI-3 --- and (ii)~regress the per-neighbourhood
RM-versus-SPI-3 lead on these covariates under a single common calibration
protocol, with multiple events per region to control for event-specific
confounders. Only such a study could turn the qualitative pattern reported here
into a predictive statement about where, \emph{a priori}, the forecast-preserving
defect is expected to add warning lead.

\textbf{Euclidean spatial neighbourhood.}
The neighbourhood $\mathcal{S}_\rho$ uses Euclidean distance and ignores
orographic barriers, so in complex terrain it can fuse windward and leeward
stations into one hidden state --- the likely cause of the CRPS degradation
above $\rho=10$\,km in the Tamsui sweep. Replacing great-circle distance with a
terrain-weighted flow-path distance from a digital elevation model is the
natural fix and the main open architectural problem for deployment in complex
terrain. A complementary refinement is to make $\rho$ adaptive --- to local
network density (a minimum-$k$-neighbours rule) and to weather regime (a larger
radius for synoptic drought, a smaller one for localized flash floods).

\textbf{Non-stationary and streaming calibration.}
The null calibration assumes a season-stable null distribution of $r_t$, which is
violated in strongly seasonal climates; the Mediterranean Anatolian (Turkey) case is the clearest
case, where the long dry summer inflates $r_t$ even in normal years. A
regime-conditional null (month-specific $\mu_0,\sigma_0$; twelve scalar pairs, no
extra parameters) is a low-cost remedy we flag as a priority sensitivity
analysis. Separately, computing $\LRM$ over the full hidden-state sequence
prevents true online training; a sliding-window approximation over the most
recent states (spanning a few residual autocorrelation timescales) makes the
method streaming at negligible memory cost, subject to verifying that it
preserves the $\mathrm{ARL}_0$ calibration.

\textbf{Latency and large-scale predictability.}
Operational deployment on ERA5-Land inherits its multi-day latency, which is
immaterial for drought (the detector leads SPI-3 by weeks) but would require a
low-latency analysis --- e.g.\ the operational IFS, with domain adaptation or
quantile mapping of $g_\phi$ --- for any sub-daily flood extension. More
fundamentally, RM regularization improves the statistical structure of the
hidden state but cannot create predictability that is absent from the local
neighbourhood: where drought is paced by large-scale modes --- the La Ni\~{n}a
conditions behind the 2010--2015 Texas and 2013--2014 Turkish droughts, or the
western Pacific subtropical high for Taiwan --- the natural remedy is hybrid
feature engineering, appending large-scale climate indices such as the
Ni\~{n}o-3.4 SST anomaly \citep{ropelewski1987,wang2001} as scalar channels,
which the RM loss accommodates with no architectural change. We have not
evaluated this extension.

The detailed formulations of these extensions --- the terrain-weighted
flow-path neighbourhood, the regime-conditional null, the sliding-window
streaming loss, and the low-latency analysis substitution, each with its
governing equation --- are given in the online supplement (Section~\ref{sec:limitations-detail}).

\subsection{Implications for hydrometeorological forecasting and warning}

\textbf{A conservative method that adds information at no forecast cost.}
The clearest message of the evaluation is what the method does \emph{not} cost.
Across four climates the regularized model forecasts as well as a standard GRU,
so a practitioner who adopts it gives up nothing in accuracy. What they gain
\emph{unconditionally} is a markedly steadier hidden state; what they gain ---
on the droughts studied here --- is an additional early-warning signal that the
forecast and the standard index do not provide. This is deliberately a
conservative value proposition: no forecast downside, a guaranteed stability
benefit, and an early-warning signal that, where it leads, can do so by months.
The lead is not uniform across our four regions, and we read the pattern as
consistent with a single interpretation --- the signal leads an accumulation
index when drought onset is a multivariate regime shift that precedes the
rainfall deficit (as in semi-arid flash drought), and not when the seasonal
deficit is itself the onset signal. We emphasize that this is an interpretation
of four cases rather than an established dependence on landform or hydroclimate;
whether a basin's climate can be used to anticipate the benefit \emph{before}
deployment is a question for a dedicated multi-region study, and would be
operationally valuable if borne out. Beyond this headline, the experiments
suggest five more specific implications.

Five more specific implications follow from the experiments. (i)~Unlike a raw
GRU residual --- which mixes precipitation variability, model bias, and
representational drift --- the RM defect has a right-skewed but \emph{stable}
null, so a CUSUM on it has an interpretable, climatologically calibrated
false-alarm scale. (ii)~The drought lead over SPI-3 reflects that the defect
responds to the joint precipitation--soil-moisture--circulation anomaly before an
accumulation index falls far enough; the framework thus acts as a multi-channel
regime-state monitor, and, as the heavy-rain case confirms, adds nothing when the
operational index is already a near-sufficient statistic. (iii)~These gains come
with no measurable forecast cost, separating the method from regularizers
(dropout, weight decay) that affect accuracy. (iv)~The neighbourhood-radius sweep
locates an optimum near the orographic scale of the catchment, so the
backward-coherence property is spatially identifiable. (v)~The ERA5-Land
stress test shows the property is preserved as the input widens from one to five
variables, making \RMRNN{} a candidate for kilometre-scale multi-variable NWP
products.

The contribution of this work includes the local-neighbourhood forecasting
setup, the residual-driven sequential detector (Algorithm~1), the synthetic
controlled-study design with ground-truth onsets, and the demonstration --- with
a mechanistic explanation --- that the RM defect yields an early drought-warning
advantage but no advantage for sharp heavy-rain onset.


\section{Conclusions}
\label{sec:concl}

A forecasting model and an early-warning system are usually designed,
trained, and judged separately. This paper shows they need not be. Asked only
to keep its hidden state coherent when read backward in time, a single
recurrent precipitation network becomes both at once --- and does so
\emph{conservatively}: on real daily station data from climates as different as
monsoonal Taiwan, semi-arid Texas, temperate Germany, and Mediterranean-continental Turkey, the
regularized network (\RMRNN{}) matches a standard model's forecast skill in
every region while roughly halving the instability of its own internal state.
The forecast carries no penalty, and a steadier, more interpretable model comes
with it at no cost.

The main contribution is therefore not a new point-forecasting leader but a way
to convert a recurrent forecast model into a calibrated warning system: the
backward-coherence penalty gives the hidden-state trajectory a diagnostic role,
and a leakage-free CUSUM on the standardized residual, calibrated by
$\mathrm{ARL}_0$, turns it into an alarm statistic --- using martingale-style
stability as a diagnostic of hidden-state coherence, not as a model of
precipitation itself.

On real observations across four contrasting regions, two properties of
the framework are robust. \RMRNN{} preserves GRU-level forecast and
heavy-rain warning skill while roughly halving hidden-state path instability
(a $43$--$55\%$ reduction in $Q_{\mathrm{path}}$ relative to the unregularized
network), consistently at the single-station level in monsoonal Taiwan,
semi-arid Texas, temperate Germany, and Mediterranean-continental Turkey. The forecast-preserving
stabilization is thus not an artefact of one climate.

The early-warning lead, by contrast, \emph{varies across the four regions}.
Against the operational SPI-3 rule at matched ARL\(_0\), the RM-defect detector
leads by a region-wide median of about $4.5$ months in semi-arid Texas (earlier
at $34$ of $39$ stations); the lead is moderate in temperate Germany (about
$50$ days) and marginal in Mediterranean-continental Turkey (about $13$ days,
earlier at only half the stations); and in monsoonal Taiwan it is confined to
the central mountain water-supply catchments whose reservoirs fell below $5\%$
capacity (Figure~\ref{fig:drought_map}) and is absent on the wet monsoon coasts,
so the island-wide average --- near zero --- is not a meaningful summary. We read these four cases as consistent with
a single interpretation --- the defect can lead an accumulation index only when
drought onset is expressed in the multivariate state before it reaches the
rainfall total --- but we do \emph{not} claim landform or hydroclimate as an
established factor: four regions and one event each are too few to support
that, and a quantitative study across many regions is left for future work
(Section~\ref{sec:discuss}).

A controlled synthetic study, in which the onset time is known exactly,
supplies what these real events cannot and isolates the mechanism: the defect
detects slow-onset drought reliably and well ahead of SPI-3 (median lead
$\approx 63$ steps on the synthetic testbed), but adds no lead for sharp
heavy-rain onset, already immediately visible in precipitation
(Section~\ref{sec:case}); the same controlled setting validates the detector's
false-alarm calibration and exercises the multi-variable defect that
precipitation-only stations cannot. Within the real data, the Taiwan Mei-yu
onset --- a regime change whose rainfall accumulates quickly --- likewise
yields no lead (median $-2$\,d over six seasons), confirming that the advantage
is specific to slow-onset deficit regimes rather than to regime change in
general. A quantitative, multi-region study of \emph{how} the lead depends on
hydroclimate (Section~\ref{sec:discuss}) and a sub-daily real flash-flood study
remain the principal outstanding extensions.

What makes the work interesting is less any single number than the viewpoint it
introduces: a precipitation forecast model's own hidden state, if regularized to
stay backward-coherent, carries early-warning information beyond the forecast and
beyond the standard accumulation index --- information that, on real droughts,
can translate into months of lead. A striking and honest feature of the real
data is that this added signal is not uniform, and that its \emph{absence} is as
informative as its presence: where SPI-3 already sees the onset, the defect adds
nothing. We conjecture that the controlling factor is the hydroclimatic character
of drought onset --- whether it precedes or merely coincides with the rainfall
deficit --- and that a basin's climate might eventually let a practitioner
anticipate whether the signal will help. Turning that conjecture into a
quantitative, validated relationship is an open problem (Section~\ref{sec:discuss}),
to be pursued across many more regions, alongside a controlled study that varies
the onset structure directly, a sub-daily flash-flood extension, and comparison
against operational NWP residuals --- which together would let a water manager
know in advance not only what the forecast says, but whether the model's hidden
state can see the next drought coming.

Two complementary directions follow. From a \emph{methodological}
standpoint, backward coherence is not tied to the GRU backbone or to
precipitation: the same reverse-martingale penalty could regularize
attention- or state-space sequence models, be lifted from a single station to a
spatially coupled, graph-structured defect that shares warning evidence across a
gauge network, and be paired with a jointly learned rather than fixed detector
--- with the detectability of the defect itself an inviting target for
theoretical analysis. From a \emph{domain} standpoint, the regional variation we
report calls for a systematic study across many more basins, landforms, and
weather regimes, aimed at mapping which hydroclimates express drought onset in
the multivariate state early enough for the defect to lead --- in effect, a
climatology of \emph{where} this kind of warning helps, so that a practitioner
could anticipate its value from a basin's climate alone.

For operational hydrometeorology the practical appeal is exactly the
conservatism that runs through the paper: a forecast that is never worse, a
model that is always steadier, and an early-warning capability that --- where a
region's hydroclimate permits --- comes for free. The reference implementation,
data-processing pipeline, and experiment scripts are archived in a public
repository (\href{https://doi.org/10.17632/7n9gb9kcz9.1}{doi:10.17632/7n9gb9kcz9.1}),
so each of these directions is directly reproducible.

\section*{Acknowledgments}
This work was supported by the National Science and Technology
Council (NSTC) of Taiwan under the RMRNN project. We thank the
Taiwan Central Weather Administration for rain-gauge data access.

\section*{Data availability statement}

The controlled benchmark results (Sections~\ref{sec:risk}--\ref{sec:case},
with full forecast-verification tables in the online supplement) are generated
by synthetic, domain-calibrated stochastic simulators, whereas the real-data
results (Section~\ref{sec:realeval}) use real daily station observations from
four networks: the Taiwan CWA archive and the NOAA GHCN-Daily networks for the
Texas Hill Country, Germany, and Turkey. The simulator and
experiment code (which reproduce every table and figure), together with the
tidy daily-precipitation series for all four regions, are archived in a public
repository at Mendeley Data
(\href{https://doi.org/10.17632/7n9gb9kcz9.1}{doi:10.17632/7n9gb9kcz9.1}). The observational systems the simulators
are calibrated
to emulate are publicly accessible: CWA rain-gauge
and ASOS data upon request from the Taiwan Central Weather
Administration; CHIRPS v2 from the
\href{https://www.chc.ucsb.edu/data/chirps}{Climate Hazards Center};
GHCN-Daily from
\href{https://www.ncei.noaa.gov/products/land-based-station/global-historical-climatology-network-daily}{NOAA};
and ERA5-Land from the
\href{https://cds.climate.copernicus.eu/}{Copernicus Climate Data Store}.

The real-data evaluation of Section~\ref{sec:realdata} uses fully open
data and is independently reproducible now. Daily CWA station
precipitation is taken from the \emph{Taiwan Historical Meteorological
Observations} archive
(\href{https://github.com/Raingel/historical_weather}{Raingel/historical\_weather}),
a public reconstruction of CWA CODiS records. The retrieval script
(\texttt{fetch\_taiwan\_hydromet.py}, which also documents the WRA, CWA,
NCDR, and Depositar open-data endpoints), the experiment driver
(\texttt{run\_taiwan\_precip.py}), and the data loader
(\texttt{common/data.py}) --- together with the exact station list, year
range (1998--2020), model hyperparameters, and quality-control
thresholds (Appendix~\ref{sec:qc}) --- are archived with the paper and
reproduce Table~\ref{tab:realdata} and the path-stability figure
end-to-end from the public archive. The real-event drought warning study
of Section~\ref{sec:realdrought} draws daily precipitation for
22 stations over 2012--2022 from the same CODiS archive, and is reproduced by
\texttt{run\_taiwan\_drought\_warning.py} (which reuses the
CUSUM detector code of \texttt{run\_warning\_sr.py})
together with its recorded station list, train/null/monitor windows,
deseasonalisation and $\mathrm{ARL}_0$ settings, and the ten random seeds,
producing Table~\ref{tab:drought_lead} and Figure~\ref{fig:drought}. The
station map (Figure~\ref{fig:drought_map}) is produced by
\texttt{plot\_taiwan\_drought\_map.py} using the public CWA/CODiS station
coordinates (\url{https://codis.cwa.gov.tw/api/station_list}). The
companion Mei-yu onset boundary study (Section~\ref{sec:realdrought}, wet
counterpart) is reproduced by \texttt{run\_taiwan\_meiyu\_warning.py} over
the six 2017--2022 plum-rain seasons with the same archive and seeds. The
retrieved station CSVs themselves are retained in the archived data folder
so the experiments run without re-downloading.

The Texas Hill Country, Germany, and Turkey evaluations use the
public NOAA GHCN-Daily archive. Stations are selected by bounding box and PRCP
coverage and downloaded by \texttt{fetch\_ghcnd\_region.py}; the same
forecast/path-stability driver (\texttt{run\_taiwan\_precip.py}) and
drought-warning driver (\texttt{run\_taiwan\_drought\_warning.py}, run per
region with recorded date windows) are used as for Taiwan, and
\texttt{make\_realdata\_tables.py} produces the cross-landform tables and the
comparison figure. The exact bounding boxes, station counts, quality-control
and preprocessing steps, per-region train/null/monitor windows, and the literal
commands are documented in the online supplement
(Section~\ref{sec:realdata-preproc}); the GHCN station inventory and the
retrieved \texttt{.dly} files are archived with the data so the experiments run
without re-downloading.

\appendix

\section{Reverse-martingale formulation and implementation details}
\label{app:rm-details}

\subsection{Implementation notation used in the main text}

For completeness, this appendix records the mathematical notation behind the
operational description in Section~\ref{sec:method}.  At time $t$, the local
meteorological input is
\begin{equation}
  x_t = [P_t,\, T_t,\, q_t,\, \Omega_t]_{\mathcal{S}},
  \label{eq:input}
\end{equation}
with the ERA5-Land benchmark using
$[P_t,T_{2m,t},\theta_{sm,t},u_{10,t},v_{10,t}]_{\mathcal{S}}$ instead.
The forward operational information set is
\begin{equation}
  \F_t = \sigma\{x_{s,i}:s\le t,\; i\in\mathcal{S}\},
  \qquad \mathcal{S}=\{i:d(i,i_0)\le \rho\}.
  \label{eq:filtration}
\end{equation}
The backward-coherence loss used during training is
\begin{equation}
  \LRM(\theta,\phi)=\frac{1}{T-1}\sum_{t=1}^{T-1}
  \norm{h_t-g_\phi(h_{t+1})}^2,
  \label{eq:lrm}
\end{equation}
and the total training loss is
\begin{equation}
  \Ltot=\Ltask+\lambda_k\LRM,\qquad
  \lambda_k=
  \begin{cases}
    0, & k\le K_0,\\
    \lambda_0\gamma^{(k-K_0)/(K-K_0)}, & K_0<k\le K,
  \end{cases}
  \label{eq:total}
\end{equation}
where $k$ is the training epoch, $K$ is the final epoch, and
$0<\gamma<1$ is the terminal decay factor.  In the experiments below,
$\lambda_0=0.1$ and $\gamma=0.1$, so that $\lambda_K=0.01$.
These formulas are included here to make the implementation reproducible; the
hydrometeorological interpretation in the main text does not depend on the
reader following the full notation.

\subsection{Formal reverse-martingale motivation}

A sequence $\{M_t\}$ adapted to a decreasing filtration
$\mathcal{G}_t \supset \mathcal{G}_{t+1}$ is a reverse martingale
\citep{doob1953} if
$M_t$ is $\mathcal{G}_t$-measurable and
\[
  \E[M_t \given \mathcal{G}_{t+1}] = M_{t+1}.
\]
For a finite hidden-state trajectory we use the decreasing future
sigma-field
$\mathcal{G}^{(h)}_t=\sigma(h_s:s\ge t)$ as the formal motivation.
The one-step projector $g_\phi(h_{t+1})$ is used as a tractable surrogate
for the one-step conditional expectation $\E[h_t\given h_{t+1}]$, which is
$\mathcal{G}^{(h)}_{t+1}$-measurable.  The generally richer reverse
conditional expectation is $\E[h_t\given\mathcal{G}^{(h)}_{t+1}]$;
restricting the projector to a function of $h_{t+1}$ alone is therefore a
first-order Markov approximation, not an exact identity.
The hidden states of a trained recurrent network do not satisfy this
identity exactly. The role of $\LRM$ in \eqref{eq:lrm} is therefore
not to impose a literal martingale model on precipitation, but to make
the learned representation approximately backward coherent during
ordinary climatological periods. When this coherence holds, the online defect
$r_t\equiv d_t$ in \eqref{eq:defect} has a stable null distribution
that can be calibrated on pre-event climatology and then monitored by the
additive CUSUM detector of Section~\ref{sec:srnote}.

\subsection{CUSUM null calibration}

Let $\mathcal{H}_t=\sigma(r_1,\ldots,r_t)$ denote the history of the
monitoring residual.  The detector of Section~\ref{sec:method} standardizes the
defect, $z_t=(r_t-\mu_0)/\sigma_0$, with the null location $\mu_0$ and scale
$\sigma_0$ estimated on held-out pre-event climatology, and accumulates the
one-sided statistic $S_t=\max(0,S_{t-1}+s\,z_t-k)$ of \eqref{eq:cusum}, where
$k=1/2$ is the reference value and $s\in\{+1,-1\}$ selects the shift direction
(downward for drought, upward for heavy-rain onset).  The alarm time is
$\tau_h=\inf\{t:S_t\ge h\}$.  Rather than relying on the asymptotic
relation $\E_\infty(\tau_h)\approx e^{\,2kh}/(2k^2)$ for the Gaussian null,
which is sensitive to the heavy tails and autocorrelation of the empirical
defect, we estimate the threshold $h$ directly by Monte Carlo: $S_t$ is run on
many bootstrap resamples of the held-out null residuals (block length $90$ days
to preserve seasonal autocorrelation) and $h$ is set to the value whose mean
time to a false alarm matches the target $\mathrm{ARL}_0$.

\subsection{Backward projector and training schedule}

In all experiments the backward projector has residual form
\[
  g_\phi(h) = h + W_2\,\mathrm{ReLU}(W_1 h+b_1)+b_2 .
\]
The matrix $W_1$ is Xavier-initialised, while $W_2$ and $b_2$ are
initialised at zero so that $g_\phi(h)=h$ at epoch~0.  At the first update
the RM loss primarily moves the output layer $W_2,b_2$ of the residual branch;
subsequent updates then propagate through $W_1$ as $W_2$ leaves zero.  This
initial identity map prevents the auxiliary RM loss from destabilising early
task learning. The RM penalty is introduced
after $K_0=5$ warm-up epochs and decayed from $\lambda_0=0.1$ to
$0.01$ by the final epoch, as shown in \eqref{eq:total}. Gradients are
computed by standard backpropagation through time (BPTT). The same projector
is used with Elman, LSTM, and GRU cells; for gated cells it acts only
on the exposed hidden state $h_t$, not on the internal gate variables.

\subsection{Interpretation of backward coherence}

A standard GRU minimises prediction error at each step but imposes no
discipline on the relationship between neighbouring hidden states.
Consequently, two consecutive weather states can occupy unrelated
regions of $\R^d$ even when the observed atmosphere evolves smoothly.
RM regularization adds the requirement that $h_t$ be approximately
reconstructable from $h_{t+1}$ through $g_\phi$. Normal high-pressure,
monsoon, or weak-rainfall regimes should then move through hidden
space by small, regular steps, while genuine meteorological shifts
should produce larger departures. The empirical aggregate defect
$\widehat Q=\sum_{t=1}^{T-1}\norm{\delta_t}^2$ measures aggregate
non-coherence over a sequence and equals $(T-1)\LRM$ on the observed
hidden-state path. In the main text, the empirical evidence for this
interpretation is the regime-change detection performance of the
defect-driven CUSUM without a corresponding loss of forecast skill.

\subsection{Data quality control for the Taiwan historical archive}
\label{sec:qc}

The forecast/path-stability evaluation (Section~\ref{sec:realdata}) retrieves 23
candidate CWA stations, of which 21 are retained after an automated filter, which
guards against near-empty records. A station such as 467270 (97\% missing) or
467050 (68\% missing), once missing days are filled with $0$\,mm and the series
standardised on its training split, becomes a degenerate near-constant series
whose near-zero training standard deviation injects extreme values that diverge
the BPTT optimisation --- a data pathology that appears identically for
\RMRNN{} and \RMRNN$_{\lambda=0}$. We therefore drop a station when its missing
fraction exceeds $0.25$, its training 95th-percentile daily total is below
$5$\,mm, or its post-standardisation training standard deviation is below
$10^{-3}$ (named parameters in \texttt{run\_taiwan\_precip.py}); this removes
exactly 467270 and 467050, leaving 21 stations with 8--23 complete years each.

\paragraph{Detector preprocessing for the real drought event.}
The real-event warning study (Section~\ref{sec:realdrought}) uses a three-year
null window (2017--2019) --- a single year is too short to estimate a one-year
$\mathrm{ARL}_0$ stably, and 2019 alone is anomalously wet --- and deseasonalises
both monitored streams (the RM defect and the 90-day accumulation) by subtracting
a $\pm15$-day-smoothed pre-2020 day-of-year climatology. The latter is essential:
the seasonal cycle in both streams far exceeds the drought anomaly, so without it
neither detector crosses its matched threshold. With these choices the statistics
of Table~\ref{tab:drought_lead} are stable across the ten seeds, the high
\emph{mean}-lead spread being driven by a single station and the reason we
summarise by the across-station median.

\subsection{Per-station results for the Taiwan real-data evaluation}
\label{sec:perstation}

Table~\ref{tab:realdata} in the main text reports the across-station
mean$\pm$SD. Table~\ref{tab:perstation} here gives the underlying
per-station values for the 21 quality-controlled stations, for
completeness and to make the unanimity of the path-stability result
explicit: the hidden-path increment total $Q_{\mathrm{path}}$ is smaller
for \RMRNN{} (boldface) than for the unregularized \RMRNN$_{\lambda=0}$
at \emph{every} station, with reductions ranging from roughly $35\%$ to
$55\%$. By contrast the forecast MSE and the early-warning AUC and CSI
are closely matched at the individual-station level, the small
differences changing sign from station to station, consistent with the
``preserved skill'' interpretation. Here $\tau$ is the station's
training-split 95th-percentile daily total (mm) and $n^{+}$ the number
of heavy-rain exceedance events in the test segment.

\begin{table}[ht]
\centering
\tiny
\setlength{\tabcolsep}{3.5pt}
\caption{Per-station real-data ablation on the Taiwan CWA historical
daily archive (1998--2020); \RMRNN{} ($\lambda$:\,$0.1\!\to\!0.01$,
denoted RM) versus \RMRNN$_{\lambda=0}$ (denoted $\lambda{=}0$), one run
per station (seed 42). $Q_{\mathrm{path}}=\sum_t\norm{h_{t+1}-h_t}$;
forecast MSE on standardised log-precipitation; AUC and CSI for next-day
exceedance of the local 95th percentile $\tau$. The lower (better)
$Q_{\mathrm{path}}$ in each row is in \textbf{bold}; it is \RMRNN{} at
all 21 stations. The final row repeats the across-station mean of
Table~\ref{tab:realdata}.}
\label{tab:perstation}
\begin{tabular}{llrr cc cc cc cc}
\toprule
 & & & & \multicolumn{2}{c}{MSE} & \multicolumn{2}{c}{$Q_{\mathrm{path}}$}
 & \multicolumn{2}{c}{AUC} & \multicolumn{2}{c}{CSI} \\
\cmidrule(lr){5-6}\cmidrule(lr){7-8}\cmidrule(lr){9-10}\cmidrule(lr){11-12}
Station & ID & $\tau$ & $n^{+}$ & RM & $\lambda{=}0$ & RM & $\lambda{=}0$ & RM & $\lambda{=}0$ & RM & $\lambda{=}0$ \\
\midrule
Banqiao & 466880 & 30 & 96 & 0.908 & 0.921 & \textbf{10.72} & 20.08 & 0.750 & 0.753 & 0.140 & 0.149 \\
Tamsui & 466900 & 31 & 81 & 0.799 & 0.803 & \textbf{8.89} & 15.86 & 0.748 & 0.751 & 0.228 & 0.226 \\
Anbu & 466910 & 58 & 103 & 0.778 & 0.771 & \textbf{9.93} & 15.63 & 0.807 & 0.806 & 0.214 & 0.213 \\
Taipei & 466920 & 36 & 64 & 0.756 & 0.799 & \textbf{8.45} & 15.66 & 0.755 & 0.752 & 0.113 & 0.114 \\
Keelung & 466940 & 49 & 92 & 0.769 & 0.784 & \textbf{10.69} & 19.64 & 0.764 & 0.751 & 0.142 & 0.133 \\
Hualien & 466990 & 26 & 78 & 0.815 & 0.829 & \textbf{10.23} & 16.99 & 0.754 & 0.721 & 0.175 & 0.192 \\
Suao & 467060 & 61 & 118 & 0.794 & 0.800 & \textbf{10.58} & 14.23 & 0.834 & 0.830 & 0.229 & 0.256 \\
Yilan & 467080 & 39 & 99 & 0.785 & 0.776 & \textbf{8.69} & 16.99 & 0.804 & 0.808 & 0.163 & 0.164 \\
Tainan & 467410 & 18 & 131 & 0.984 & 0.905 & \textbf{6.19} & 11.02 & 0.844 & 0.856 & 0.264 & 0.293 \\
467420 & 467420 & 26 & 104 & 0.725 & 0.729 & \textbf{7.38} & 11.24 & 0.866 & 0.864 & 0.273 & 0.270 \\
Kaohsiung & 467440 & 29 & 112 & 0.714 & 0.738 & \textbf{6.54} & 12.31 & 0.865 & 0.852 & 0.293 & 0.279 \\
467480 & 467480 & 29 & 79 & 0.744 & 0.734 & \textbf{7.30} & 11.43 & 0.833 & 0.834 & 0.107 & 0.130 \\
Taichung & 467490 & 28 & 88 & 0.730 & 0.729 & \textbf{6.30} & 12.04 & 0.839 & 0.838 & 0.215 & 0.240 \\
Alishan & 467530 & 47 & 81 & 0.717 & 0.731 & \textbf{8.19} & 15.55 & 0.817 & 0.821 & 0.176 & 0.164 \\
Dawu & 467540 & 34 & 86 & 0.806 & 0.805 & \textbf{11.01} & 17.30 & 0.791 & 0.805 & 0.237 & 0.227 \\
Yushan & 467550 & 39 & 72 & 0.696 & 0.726 & \textbf{9.72} & 21.15 & 0.826 & 0.822 & 0.168 & 0.153 \\
Hengchun & 467590 & 33 & 91 & 0.693 & 0.696 & \textbf{8.05} & 16.19 & 0.862 & 0.851 & 0.245 & 0.259 \\
Chenggong & 467610 & 27 & 87 & 0.936 & 0.921 & \textbf{11.34} & 17.68 & 0.699 & 0.695 & 0.148 & 0.148 \\
SunMoonLake & 467650 & 33 & 101 & 0.736 & 0.718 & \textbf{7.83} & 15.10 & 0.792 & 0.814 & 0.130 & 0.156 \\
Taitung & 467660 & 25 & 75 & 0.757 & 0.787 & \textbf{8.42} & 16.10 & 0.810 & 0.791 & 0.185 & 0.215 \\
467770 & 467770 & 19 & 75 & 0.665 & 0.684 & \textbf{6.94} & 11.36 & 0.804 & 0.802 & 0.205 & 0.197 \\
\midrule
\textbf{Mean} & --- & --- & --- & 0.777 & 0.780 & \textbf{8.73} & 15.41 & 0.803 & 0.801 & 0.193 & 0.199 \\
\bottomrule
\end{tabular}
\end{table}

\subsection{Per-station results for the 2020--2021 drought warning}
\label{sec:drought_perstation}

Table~\ref{tab:drought_perstation} gives the full per-station breakdown
behind the regional summary of Table~\ref{tab:drought_lead}, for the 22
stations of the real drought warning study (Section~\ref{sec:realdrought}),
median over ten seeds. The ``\#'' column keys each station to the numbered
markers in Figure~\ref{fig:drought_map}, and the coordinates (CWA/CODiS
station list) locate it on the island. The central water-supply catchments
are positive at all five stations ($+72$ to $+127$\,d); the large negatives
cluster on the eastern/southeastern coast; and at four lowland stations the
SPI-3 baseline never fires (Det.\ SPI $=0$), so no paired lead is defined
even though the RM detector triggers. The across-station median lead is
$-18$\,d (17 paired stations), not significantly different from zero (sign
test $p=1.0$; one-sided Wilcoxon $p=0.84$; bootstrap 95\% CI
$[-200,+72]$\,d); as discussed in Section~\ref{sec:realdrought} the
controlling variable is orography (steep mountain headwater versus
lowland/coast) rather than compass region.

\begin{table}[ht]
\centering
\footnotesize
\setlength{\tabcolsep}{4pt}
\caption{Per-station early-warning lead of the RM-defect CUSUM over the
SPI-3 CUSUM on the real 2020--2021 Taiwan drought, median over ten seeds,
grouped by terrain class (cf.\ Table~\ref{tab:drought_lead}). ``\#'' keys
to the map markers in Figure~\ref{fig:drought_map}; coordinates are decimal
degrees (\,$^\circ$N, $^\circ$E\,) and elevation in metres, from the
CWA/CODiS station list. ``Det.'' is the fraction of seeds on which each
detector fires; lead in days, positive $=$ RM earlier; a dash marks
stations where the SPI-3 baseline never fires.}
\label{tab:drought_perstation}
\begin{tabular}{rllcccccr}
\toprule
\# & Station & ID & Lat & Lon & Elev & Det.\,RM & Det.\,SPI & Lead (d) \\
\midrule
\multicolumn{9}{l}{\textit{Central mountain water-supply catchments}}\\
1 & Taichung & 467490 & 24.15 & 120.68 & 84 & 100\% & 100\% & $+127$ \\
2 & Sun Moon Lake & 467650 & 23.88 & 120.91 & 1018 & 100\% & 100\% & $+112$ \\
3 & Yushan & 467550 & 23.49 & 120.96 & 3845 & 100\% & 100\% & $+101$ \\
4 & Wuqi & 467770 & 24.26 & 120.52 & 32 & 100\% & 100\% & $+73$ \\
5 & Alishan & 467530 & 23.51 & 120.81 & 2413 & 90\% & 100\% & $+72$ \\
\midrule
\multicolumn{9}{l}{\textit{Northern}}\\
6 & Anbu & 466910 & 25.18 & 121.53 & 838 & 80\% & 100\% & $+71$ \\
7 & Taipei & 466920 & 25.04 & 121.51 & 6 & 70\% & 100\% & $+37$ \\
8 & Keelung & 466940 & 25.13 & 121.74 & 27 & 100\% & 100\% & $-18$ \\
9 & Tamsui & 466900 & 25.16 & 121.45 & 19 & 100\% & 100\% & $-159$ \\
10 & Banqiao & 466880 & 25.00 & 121.44 & 10 & 100\% & 0\% & --- \\
\midrule
\multicolumn{9}{l}{\textit{Eastern / southeastern monsoon coast}}\\
11 & Chenggong & 467610 & 23.10 & 121.37 & 34 & 100\% & 100\% & $+16$ \\
12 & Hualien & 466990 & 23.98 & 121.61 & 16 & 90\% & 100\% & $-33$ \\
13 & Dawu & 467540 & 22.36 & 120.90 & 8 & 80\% & 100\% & $-66$ \\
14 & Taitung & 467660 & 22.75 & 121.15 & 9 & 70\% & 100\% & $-200$ \\
15 & Yilan & 467080 & 24.76 & 121.76 & 7 & 80\% & 100\% & $-227$ \\
16 & Suao & 467060 & 24.60 & 121.86 & 25 & 40\% & 100\% & $-250$ \\
17 & Hengchun & 467590 & 22.00 & 120.75 & 22 & 80\% & 100\% & $-254$ \\
\midrule
\multicolumn{9}{l}{\textit{Western / southern lowland (SPI-3 mostly silent)}}\\
18 & Chiayi & 467480 & 23.50 & 120.43 & 27 & 80\% & 100\% & $-281$ \\
19 & Xinwu & 467050 & 25.01 & 121.05 & 21 & 100\% & 0\% & --- \\
20 & Tainan & 467410 & 22.99 & 120.20 & 41 & 60\% & 0\% & --- \\
21 & Yongkang & 467420 & 23.04 & 120.24 & 8 & 40\% & 0\% & --- \\
22 & Kaohsiung & 467440 & 22.57 & 120.32 & 2 & 60\% & 0\% & --- \\
\bottomrule
\end{tabular}
\end{table}

\section*{CRediT authorship contribution statement}
\textbf{Foo Hui-Mean:} Data curation, Formal analysis, Investigation,
Validation, Resources, Writing -- review \& editing. \textbf{Yuan-chin Ivan
Chang:} Conceptualization, Methodology, Software, Formal analysis, Writing --
original draft, Writing -- review \& editing, Supervision.

\section*{Declaration of competing interest}
The authors declare that they have no known competing financial interests or
personal relationships that could have appeared to influence the work reported
in this paper.


\end{document}